\documentclass[preprint]{aastex63_mod}

 \usepackage{amssymb}
 \usepackage{amsmath}
 \usepackage{amsmath,bm}
 \usepackage{graphicx}
\usepackage{savesym}
\savesymbol{tablenum}
\usepackage{siunitx}
\restoresymbol{SIX}{tablenum}
\usepackage{rotating}
\newcommand{\iu}{\mathrm{i}}
\newcommand{\ex}{\mathrm{e}}

\newcommand{\matrixB}[1]{\bm{\mathrm{#1}}}

\usepackage{mathtools}
\usepackage[thinc]{esdiff}

\usepackage{changes}
\newcommand{\stkout}[1]{\ifmmode\text{\sout{\ensuremath{#1}}}\else\sout{#1}\fi}
\setdeletedmarkup{\stkout{#1}}
\accepted{27 March 2024}
\published{in PSJ}

\shorttitle{A Spectral Method for 3D Tides}
\shortauthors{Rovira et al.}
\graphicspath{{./}{Figures/}}

\begin{document}

\title{A Spectral Method to Compute the Tides of Laterally-Heterogeneous Bodies}

\author[0000-0002-9980-5065]{Marc Rovira-Navarro}
\affiliation{Lunar and Planetary Laboratory, University of Arizona, Tucson, AZ 85721, USA}
\affiliation{Faculty of Aerospace Engineering, TU Delft, Building 62 Kluyverweg 1, 2629 HS Delft, the Netherlands}

\author{Isamu Matsuyama}
\affiliation{Lunar and Planetary Laboratory, University of Arizona, Tucson, AZ 85721, USA}

\author{Alexander Berne}
\affiliation{California Institute of Technology, Pasadena, CA 91125}



\begin{abstract}

Body tides reveal information about planetary interiors and affect their evolution. Most models to compute body tides rely on the assumption of a spherically-symmetric interior. However, several processes can lead to lateral variations of interior properties. We present a new spectral method to compute the tidal response of laterally-heterogeneous bodies. Compared to  previous spectral methods, our approach is not limited to small-amplitude lateral variations; compared to finite element codes, the approach is more computationally-efficient. While the tidal response of a spherically-symmetric body has the same wave-length as the tidal force; lateral heterogeneities produce an additional tidal response with an spectra that depends on the spatial pattern of such variations. For Mercury, the Moon and Io the amplitude of this signal is as high as $1\%-10\%$ the main tidal response for long-wavelength shear modulus variations higher than $\sim 10\%$ the mean shear modulus. For Europa, Ganymede and Enceladus, shell-thickness variations of $50\%$ the mean shell thickness can cause an additional signal of $\sim 1\%$ and $\sim 10\%$ for the Jovian moons and Encelaudus, respectively. Future missions, such as \textit{BepiColombo} and \textit{JUICE}, might measure these signals. Lateral variations of viscosity affect the distribution of tidal heating. This can drive the thermal evolution of tidally-active bodies and affect the distribution of active regions.

\end{abstract}

\keywords{Solid body tides --- Planetary Interior  --- Natural satellites (Solar system)}

\section{Introduction}
The deformation of planets and moons under a tidal force depends on their internal properties ---e.g., density stratification, elastic and anelastic properties. As a consequence, observations of body tides can be used to infer the internal properties of planetary objects \citep[e.g.,][]{Konopliv1996,Wahr2006,Steinbr2018,Williams2015,Sohl2014,Hoppa1999,SEGATZ1988,HAMILTON2013}. 

Traditionally, body tides have been studied under the assumption of spherical symmetry, i.e., the internal properties of the body do only vary radially and not laterally. However, planets and moons are not spherically symmetric. Seismic tomography evidences that Earth's lithosphere and mantle are not spherically symmetric but present lateral structures associated with tectonics and deep mantle dynamics \citep{Woodhouse1984,Ritsema2011}; the Moon shows a farside-nearside dichotomy in crustal thickness \citep{Neumann1996,wieczorek2013}; Mars has a well-known Northern-Southern hemisphere dichotomy \citep{Watters2007,Neumann2004} and so does the ocean world Enceladus  \citep{Porco2006,Crow-Willard2015,Beuthe2016,Cadek2016,Park2024}.

Lateral variations of internal properties affect the tidal response of a body. If a body is spherically symmetric, the tidal response is proportional to the tidal forcing --- the body responds with a mode of same wavelength as that of the forcing. In contrast, when internal properties vary laterally, modes of different wavelength are coupled and the tidal response contains modes of wavelengths other than that of the forcing \citep[e.g.,][]{Berne2023b}. The tidal response spectra contains information about the amplitude and patterns of lateral variations, hence, it can be used to constrain such variations; a technique known as tidal tomography \citep[e.g.,][]{Métivier2007,Qin2014,Qin2016,Zhong2012,Lau2017}. Lateral variations of anelastic properties affect the distribution of tidal energy dissipation within the interior \citep{STEINKE2020}, affecting the evolution of the interior and resulting in distinct surface features.

The tidal response of a spherically-symmetric body is often obtained by solving the equations governing the deformation of a self-gravitating body using a spectral method with spherical harmonics as basis functions. The properties of spherical harmonics allow to transform the $3D$ governing differential equations to a set of equations with radial dependence only. \cite{Love1911} first  used this approach to obtain the tidal response of a self-gravitating  homogeneous elastic body. Since then, the approach of Love has been extended to radially-stratified bodies \citep[e.g.,][]{Takeuchi1962,Alterman1959,JaraOrue2011} and to include anelastic effects \citep[e.g.,][]{Peltier1974,Wahr1986,Ross1986,SEGATZ1988}. This approach is computationally inexpensive and allows to tackle the inverse problem ---given the tidal response of a body, infer its interior properties.

Mode-coupling makes it more complicated to obtain the tidal response of a body with lateral heterogeneities. Several methods have been developed to tackle the problem. Finite element methods (FEM) have been employed for tidal \citep[e.g.,][]{A2014,Métivier2006,LATYCHEV200986,STEINKE2020,Zhong2012,SOUCEK2017,SOUCEK2019,zhong2022citcomsve,Berne2023a} and surface loads \citep[e.g.,][]{Wu2004,vanderWal2013}. Alternatively, spectral methods can be used. Martinec and collaborators have developed a numerical spectral-finite element approach to obtain the response of a self-gravitating viscoelastic Earth to surface loads \citep{Martinec2000,Tanaka2011,Tanaka2021}. In their approach, the angular dependence of the solution is expanded using tensor spherical harmonics and the resulting equations solved using finite elements in the radial direction. Numerical methods are very flexible but they are computationally expensive, making them unsuitable to solve the inverse problem. To overcome this limitation, several authors have developed inexpensive semi-analytical spectral methods. However, existing methods either rely on perturbation theory and thus are only applicable to bodies with  small lateral heterogeneities \citep{Qin2014,Qin2016,Lau2015,Lau2017}, or on thin-shell theory and are therefore limited to bodies that have a thin outer shell overlying a fluid layer \citep{BEUTHE2018145,BEUTHE2019}.

The aim of this work is to present a new spectral method that combines the flexibility of numerical methods with the computational efficiency of semi-analytical spectral methods and apply it to consider the tidal response of bodies with lateral heterogeneities. As opposed to spectral-perturbation methods, our method does not require lateral variations to have a small amplitude. Compared to the spectral-finite-element method of \cite{Martinec2000,Tanaka2021}, angular integrals are performed analytically using the properties of tensor-spherical harmonics and not numerically; and the equations are solved in the Fourier domain, rather than in the time domain (as done for lateral variations of viscosity in \cite{Martinec2000}) or iteratively (as done for lateral variations of elastic properties in \cite{Tanaka2021}). 

The paper is structured as follows. Section \ref{sec:problem-formulation} introduces the viscoelastic-gravitational problem, in Sections \ref{sec:spectral-approach}  and \ref{sec:numerical-method} the governing equations are transformed to the spectral domain and solved. In Section \ref{sec:model-benchmark} we compare the new spectral method with a spectral-perturbation method and a FEM code; and in Sections \ref{sec:elastic-body} and \ref{sec:viscoelastic} we obtain the tidal response of elastic and viscoelatic bodies to illustrate with various types of lateral heterogeneities. The paper finishes with a summary of the main results in Section \ref{sec:discussion-and-conclusions}. 

Alongside with this paper, we release the \textit{LOV3D} software package, which uses the method described here to compute the tidal response of laterally-heterogeneous bodies \citep{LOV3D}. 
\section{Problem Formulation}
\label{sec:problem-formulation}

\subsection{Governing Equations}
The tidal response of a self-gravitating body can be obtained by solving the mass, momentum and Poisson's equation. The equations are linearised around a state of hydrostatic equilibrium with gravitational potential $\phi_0$ and pressure $p_0$ given by $\nabla p_0=-\rho_0\nabla\phi_0$ and $\nabla^2\phi_0=4\pi G\rho_0$, where $G$ is the universal gravitational constant and $\rho_0$ the density of the unperturbed body. The resulting equations are then \citep[e.g.,][]{Sabadini2016}: 

\begin{subequations}
\begin{equation}
    \rho'=-\rho_0\chi-\bm{u}\cdot\nabla\rho_0,
\end{equation}
\begin{equation}
    \nabla\cdot\bm\sigma'-\rho_0\nabla(g\bm{u}\cdot \bm{e_r})+g\rho_0\chi\bm{e_r}-\rho_0\nabla\phi'=0,
    \label{eq:momentum-0}
\end{equation}
\begin{equation}
    \nabla^2\phi'=4\pi G\rho'.
    \label{eq:Poisson-0}
\end{equation}
\label{eq:equations-of-motion}
\end{subequations}
$\bm{u}$ is the displacement vector. $\bm\sigma'$ is the incremental material stress tensor, and $\rho'$ and  $\phi'$ are the incremental local density and gravitational potential, respectively ---with the latest including both the tidal potential and the potential arising from the deformation of the body. $\chi=\nabla\cdot\bm{u}$ is the divergence of the displacement vector, $\bm{e}_r$ is the radial vector and $g$ the gravity of the unperturbed body.

To close the system, a constitutive equation relating stress and displacements is required. For an elastic body, we have
\begin{equation}
    \bm\sigma'=\lambda\chi\bm{I}+\mu\left(\nabla\bm{u}+(\nabla\bm{u})^\dag\right)
    \label{eq:rheology}
\end{equation}
where $\dag$ indicates transpose and $\bm{I}$ is the identity matrix. $\lambda$ is the second Lame parameter ($\lambda=\kappa-\frac{2}{3}\mu$), and $\mu$ and $\kappa$ are the shear and bulk modulus. If the body is viscoelastic, the correspondence principle can be used to relate stress and displacement \citep[e.g.,][]{Peltier1974,Sabadini2016}. Eq. (\ref{eq:rheology}) still holds but now $\mu$ and $\lambda$ are the Fourier-transformed shear modulus and Lame parameter $\hat\mu$, $\hat\lambda$. For Maxwellian rheology, we have: 
\begin{subequations}
\begin{equation}
    \hat\mu^{-1}=\frac{1}{\mu}\left[1-\frac{\iu}{\omega\tau_M}\right]
\end{equation}
\begin{equation}
    \hat\lambda=\kappa-\frac{2}{3}\hat\mu
\end{equation}
\label{eq:correspondance-principle}
\end{subequations}
$\omega$ is the forcing frequency and $\tau_M$ is the Maxwell time $\tau_M=\eta/\mu$, with $\eta$ being the viscosity. Different expressions for the Fourier-transformed shear modulus and Lame parameter can be obtained for other rheological models (e.g., Andrade, Voigt-Kelvin, Burgers, Sundberg-Cooper)  \citep{Renaud2018}.  

The tidal response is obtained by solving Eqs.~(\ref{eq:equations-of-motion},\ref{eq:rheology}) for a given tidal load, interior structure and boundary conditions.  

\subsection{Interior Structure and Model Assumptions}
\label{sec:non-dimensional-numbers}
We consider a layered-interior. For each layer the interior properties remain constant with radius. Each layer is characterized by its outer radius, $R_i$, density, $\rho_i$ and its mechanical properties. The mechanical properties (i.e, shear modulus, bulk modulus and viscosity) are written as a mean value ($\mu_0$, $\kappa_0$ and $\eta_0$) and a lateral variation 
\begin{equation}
\left(\mu(r,\theta,\varphi),\kappa(r,\theta,\varphi),\eta(r,\theta,\varphi)\right)=(\mu_0(r),\kappa_0(r),\eta_0(r))+\sum_{n\neq0,m}(\mu_{0}\mu_{n}^m(r),\kappa_{0}\kappa_{n}^m(r),\eta_{0}\eta_{n}^m(r))Y_n^m(\theta,\varphi).
\label{eq:rheology-expanded}
\end{equation}

$Y_n^m$ are spherical harmonics of degree $n$ and order $m$ (see Appendix \ref{ap:rank-0-tensor}), and $\theta$ and $\varphi$ are the co-latitude and longitude, respectively.

As in previous work \citep[e.g.,][]{Qin2016}, we account for the effect of thickness variations $\Delta H$ of a layer using using effective shear modulus variations: $\Delta\mu/\mu_0=\Delta H/H_0$ \citep[e.g.,][]{Qin2016}. Yet this is an approximation that deserves further comment. The tidal response of a body whose outer layer is a thin solid shell floating above a liquid layer (i.e., thin-shell body) is controlled by the extension and the bending rigidity: $2(1+\nu)\mu H$ and $\mu H^3/6(1-\nu)$ \citep{BEUTHE2018145}. The effect of shear modulus and shell thickness variations in the extension rigidity is equivalent. However, this is not true for the bending rigidity. Bending effects are small provided $(H/R)^2\ll 1$ and the response has a characteristic long wavelength; when this is the case, they can be neglected (membrane approximation). As bending effects become more relevant (e.g., thicker outer shells and shorter wavelength), the use of an effective shear modulus to mimic thickness variations becomes less accurate. \cite{BEUTHE2018145} applied thin shell theory to study the deformation of Enceladus with lateral shell thickness variations and showed that the membrane approximation renders good results provided shell variations are of long wavelength. As bending effects are expected to be smaller in larger icy moons (e.g., Europa and Ganymede), we expect this approximation to hold better for these moons. 

While the method described below can be applied to any layered interior structure, we will consider a simplified 3-layer model consisting of a non-deformable solid core of density $\rho_1$ surrounded by an incomprehensible liquid layer in hydrostatic equilibrium with outer radius $R_2$ and density $\rho_2$; and a solid shell with outer radius $R$, density $\rho_3$ and mean shear modulus, bulk modulus and viscosity $\mu_{0}$, $\kappa_{0}$ and $\eta_0$. This simple model can be used to represent rocky and icy bodies. For the former, the innermost two layers correspond to the inner and the outer core and the outermost layer represents both the rocky mantle and crust. For the latter, the core corresponds to the rocky core, the liquid layer to a subsurface ocean and the outer layer to the ice shell. This simplified model neglects the gravitational coupling between the inner solid core and the surrounding shell, which typically have a small effect on the tidal response, as well as dynamic liquid tides in the liquid layer, which under certain circumstances can play a relevant role in the tidal response \citep[e.g.,][]{RoviraNavarro2023,ROVIRANAVARRO2019}.

The tidal response is controlled by a set of non-dimensional parameters. The tidal response of an incompressible homogeneous elastic body only depends on the effective rigidity $\mu_{\textrm{eff}}$. A body with high effective rigidity exhibits small tidal deformations. The role of compressibility depends on the Poisson's ratio $\nu$. For $\nu\approx0.5$, the body behaves as an incompressible body. Viscous effects depend on the non-dimensional Maxwell time $\tau$. Bodies with $\tau\rightarrow\infty$ behave elastically. The introduction of a solid core and a liquid layer adds three non-dimensional parameters: the ratio between the outer radius of the liquid layer and surface radius, $r_R$; the ratio between the outer layer density and the mean density, $r_\rho$; and the density contrast between the liquid and solid shells ,$r_{\Delta\rho}$. For icy moons with subsurface oceans $r_{\Delta\rho}\approx 0$, while for rocky worlds we will assume that the inner and outer core have the same density. Lateral rheology variations introduce additional non-dimensional parameters. The effect of lateral variations of rheology properties depend on the parameters $\kappa_n^m$, $\mu_n^m$ and $\eta_n^m$. Characteristic non-dimensional parameters for tidally-active Solar System bodies are listed in Table \ref{tab:non-dimensional-numbers}.

\begin{table}
\begin{center}
\begin{tabular}{ c c c c c c c c}
 \hline
Parameter & Definition & Moon & Mercury &  Io & Europa & Ganymede & Enceladus \\
\hline
$\mu_{\textrm{eff}}$ & $\mu_{0}/g_s\bar\rho R$ & 
$6.38$ & 
$1.22$  & 
$5.20$  & 
$0.54$ &
$0.45$ &
$71.70$ \\ 
$r_R$  & $R_{2}/R$ & 
$0.22$ & 
$0.80$  & 
$0.53$  & 
$0.98$ &  
$0.95$ &
$0.91$  \\ 
$r_\rho$ & $\rho_3/\bar{\rho}$ & 
$0.98$ & 
$ 0.54$  & 
$0.92$  & 
$0.33$ & 
$0.52$ &
$0.62$  \\ 
$r_{\Delta\rho}$ & $(\rho_2-\rho_3)/\bar{\rho}$ & 
$1.60$ & 
$0.65$  & 
$0.54$  & 
$0$ & 
$0$ &
$0$  \\ 
$\nu$  & $\frac{3\kappa_0-2\mu_0}{6\kappa_0+2\mu_0}$  & 
$0.36$ & 
$0.36$  & 
$0.36$  & 
$0.33$  & 
$0.33$ &
$0.33$ \\ 
$\tau$  & $\omega\eta_{0}/\mu_{0}$ &  
$-$ & 
$-$  & 
$3.42$  & 
$-$  & 
$0$ &
$-$ \\ 
$\mu_n^m$, $\Delta\mu/\mu_0$ ($\Delta H/H_0$)  & Eq.~(\ref{eq:rheology-expanded}) & 
$-$ & 
$-$  & 
$0.1$  & 
$0.2$  & 
$-$ &
$1$ \\ 
$\kappa_n^m$, $\Delta\kappa/\kappa_0$  & Eq.~(\ref{eq:rheology-expanded}) & 
$-$ & 
$-$  & 
$-$  & 
$-$  & 
$-$ &
$-$ \\ 
$\eta_n^m$, $\Delta\eta/\eta_0$  & Eq.~(\ref{eq:rheology-expanded}) & 
$-$ & 
$-$  & 
$1$  & 
$-$  & 
$-$ &
$-$ \\ 
$k_2^u$  & Eq.~(\ref{eq:Love-aspherical}) & 
$0.025$ & 
$0.44$  & 
$0.056-0.0150\iu$  & 
$0.23$  & 
$0.4$ & 
$ 0.016$ \\ 
$h_2^u$  & Eq.~(\ref{eq:Love-aspherical}) & 
$0.042$ & 
$0.8$  & 
$0.1-0.026\iu$  & 
$1.16$  & 
$1.3$ & 
$0.044$ \\ 
 \hline
\end{tabular}
\end{center}
\caption{Representative values for the non-dimensional parameters for some Solar System bodies. The shear and bulk modulus of ice and rock are from \cite{SOUCEK2019} and \cite{Kervazo2021}, respectively. For the moon, we consider a core radius of $390$ \si{km} and adjust the core and mantle densities to meet the density density and $MoI$ constraint. For Mercury the interior structure follows form the average density, and the total and outer shell moment of inertia \citep{Genova2019}. Io's interior model is from \cite{STEINKE2020}. Ice shell thickness for Europa, Ganymede and Enceladus are from \cite{Beuthe2016}, \cite{Hussmann2002} and \cite{Vance2018}; we assume ice and ocean densities of $1000$ \si{kg.m^{-3}} and the core density is set to meet the mean density constraint. Io's viscosity is adjusted to match the $\Im({k_2})$ from astrometric observations \citep{Lainey2009}. Lateral variations are given in terms of maximum peak-to-peak variations with respect to the mean value. For Io, representative viscosity and shear modulus variations follow from expected variations in tidal heating \citep{Steinke_thesis}; for Enceladus and Europa expected
shell thickness variations are from \citep{Beuthe2016} and \cite{NIMMO2007}, respectively. The gravitational and radial displacement Love numbers ($k_2^u$ and $h_2^u$) for the spherically symmetric elastic case is also given.}
\label{tab:non-dimensional-numbers}
\end{table}

\section{Spectral Method}
\label{sec:spectral-approach}
Different approaches can be used to solve Equation (\ref{eq:equations-of-motion}). We use a spectral method that employs tensor spherical harmonics as basis functions \citep{James1976}. The properties of spherical harmonics allows to transform the three-dimensional governing equations to a set of partial differential equations that depends only on radial distance. In this section, we describe the new spectral method. First, we explain how the different fields are expanded using tensor spherical harmonics (Section \ref{subsec:spectral-expansion}); we then provide the governing equations and boundary conditions in the spectral domain (Sections \ref{sec:spectral-form-equations-motion} and \ref{sec:boundary-condition}); finally, we extend the definition of Love numbers traditionally used for spherically-symmetric bodies to laterally-heterogeneous bodies and obtain expressions to compute tidal energy dissipation for the anelastic case (Sections \ref{sec:Love-numbers-aspherical-body} and \ref{sec:energy-dissipation}). The appendices contain further details about the method. Appendix \ref{ap:tensor-spherical-harmonics} provides definitions of tensor spherical harmonics; Appendix \ref{ap:coupling-coefficients} gives explicit expressions to compute various integrals of products of tensor spherical harmonics required for this approach; and Appendix \ref{ap:explicit-equations} provides explicit expressions for the governing equations. 

\subsection{Spectral Expansion}
\label{subsec:spectral-expansion}
 As we are considering a periodic forcing, it is convenient to work in the Fourier domain. The tidal potential, $\phi^T$, can be written as
\begin{equation}
    \phi^T(r,\theta,\varphi,t)=\sum_{k}\left(\frac{r}{R}\right)^{n_k}{\phi^{(T)}}_{n_k}^{m_k,\omega_k}Y_{n_k}^{m_k}(\theta,\varphi)\ex^{\iu\omega_k t}.
    \label{eq:generic-forcing}
\end{equation}
$n_k$, $m_k$ and $\omega_k$ are respectively the degree, order and frequency of the tidal force. $\phi^T$ should be real, implying that for each $n_k$, $m_k$, $+\omega$ component there is a $n_k$, $-m_k$ $-\omega$ component  with amplitude ${\phi^{(T)}}_{n_k}^{-m_k,-\omega_k}=(-1)^m\overline{{\phi^{(T)}}_{n_k}^{m_k,\omega_k}}$, where $\overline{\phantom{A}}$ indicates complex-conjugate. The amplitude of ${\phi^{(T)}}_{n_k,\omega_k}^{m_k}$ depends on the tidal component considered ---Appendix \ref{sec:tidal-potential-synch-sat} provides the expression corresponding to a synchronous satellite in an eccentric orbit. As Eq.~(\ref{eq:equations-of-motion}) is linear, the complete tidal response can be obtained by linearly adding the solution for each tidal component. In what follows, we will consider the solution to a tidal force of frequency $\omega$ of the form
\begin{equation}
    \phi^T(r,\theta,\varphi,t)=\left(\frac{r}{R}\right)^{n_T}\Big({\phi^{(T)}}_{n_T}^{m_T,+}Y_{n_T}^{m_T}(\theta,\varphi)\ex^{\iu\omega t}+(-1)^{m_T}\overline{{\phi^{(T)}}_{n_T}^{m_T,+}}Y_{n_T}^{-m_T}(\theta,\varphi)\ex^{-\iu\omega t}\Big).
    \label{eq:generic-forcing-2}
\end{equation}
The $+$ and $-$ components correspond to the $+\omega$ and $-\omega$ components of the tidal force.

We expand rank $0$ tensors, such as the perturbing potential $\phi'$ and the divergence of the displacement vector $\nabla\cdot\bm{u}$, using spherical harmonics of rank $0$
\begin{equation}    (\phi'(r,\theta,\varphi,t),\chi(r,\theta,\varphi,t))=\sum_{n,m}\left[(\phi_n^{m,+}(r),\chi_n^{m,+}(r))\ex^{\iu\omega t}+(\phi_n^{m,-}(r),\chi_n^{m,-}(r))\ex^{-\iu\omega t}\right]Y_n^m(\theta,\varphi).
    \label{eq:rank-0-expansion}
\end{equation}
where summation is over $n\in[0,\infty]$ and $m\in[-n,n]$. 

The rheology parameters are also expanded in spherical harmonics (Eq.~\ref{eq:rheology-expanded}). For an anelastic body, the complex shear modulus  is obtained using the correspondence principle (Eq.~(\ref{eq:correspondance-principle}))
\begin{equation}
    \hat{\mu}(r,\theta,\varphi)=\mu_{0}(r)(1+\sum_{n\neq0,m}\mu_{n}^m(r)Y_n^m(\theta,\varphi))\left(1-\frac{\iu}{\omega\tau_{M,0}(r)}\frac{1+\sum_{n\neq0,m}\mu_{n}^m(r)Y_n^m(\theta,\varphi)}{1+\sum_{n\neq0,m}\eta_{n}^m(r)Y_n^m(\theta,\varphi)}\right)^{-1}
    \label{eq:complex-shear-1}
\end{equation}
with $\tau_{M,0}=\eta_{0}/\mu_{0}$ the average Maxwell time. The Fourier-transformed shear modulus can be decomposed into spherical harmonics
\begin{equation}
    \hat{\mu}(r,\theta,\varphi)=\hat\mu_0(r)+\mu_0\sum_{n\neq0,m}\hat\mu_n^m(r)Y_n^m(\theta,\varphi)
\end{equation}
with
\begin{equation} 
\left(\Re(\hat\mu_n^m),\Im(\hat\mu_{n}^m)\right)=\frac{1}{4\pi\mu_0}\int_\Omega(\Re(\hat\mu),\Im(\hat\mu))\overline{Y_n^{m}}\textrm{d}\Omega.
    \label{eq:complex-shear-2}
\end{equation}
$\Re$ and $\Im$ denote the real and imaginary components, respectively. For an elastic body we have $\hat\mu_0=\mu_0$ and $\hat{\mu}_n^m=(-1)^{m}\overline{\hat{\mu}_n^{-m}}$.

The displacement vector is expanded using tensor spherical harmonics of rank $1$ (Appendix \ref{ap:rank-1-tensor-spherical-harmonics})
\begin{equation}
     \bm{u}(r,\theta,\varphi,t)=\sum_{n,n_1,m}\left[u_{n,n_1}^{m,+}(r)\ex^{\iu\omega t}+u_{n,n_1}^{m,-}(r)\ex^{-\iu\omega t}\right]\bm{Y}^m_{n,n_1}(\theta,\varphi)
    \label{eq:rank-1-expansion}
\end{equation}
where summation is over $n\in[0,N]$,  $m\in[-n,n]$ and $n_1\in[n-1,n+1]$. The decomposition in tensor spherical harmonics automatically separates between spheroidal ($n_1=n\pm 1$) and  toroidal components $n_1=n$. This expansion is different to that traditionally used to solve the viscoelastic-gravitational problem, which uses the radial–poloidal–toroidal decomposition
\begin{equation}    \bm{u}(r,\theta,\varphi,t)=\sum_{n,m}\left(U_n^{m,+}(r)\bm{R}_n^m(\theta,\varphi)+V_n^{m,+}(r)\bm{S}_n^m(\theta,\varphi)+W_n^{m,+}(r)\bm T_n^m(\theta,\varphi)\right)\ex^{\iu\omega t}+c.c,
\end{equation}
where $c.c$ means complex conjugate. $U_n^m,V_n^m,W_n^m$ and $u_{n,n-1}^m$,  $u_{n,n}^m$, $u_{n,n+1}^m$ are related via a linear transformation (Eq.~\ref{eq:displacement-vector-rad-tan}). 

The stress and strain tensors are expanded using tensor spherical harmonics of rank $2$ (Appendix \ref{ap:rank-2-tensor})
\begin{equation}    \left(\bm\sigma'(r,\theta,\varphi,t),\bm\epsilon'(r,\theta,\varphi,t)\right)=\sum_{\substack{n,n_1,\\n_2,m}}\left[\left(\sigma_{n,n_1,n_2}^{m,+}(r),\epsilon_{n,n_1,n_2}^{m,+}(r)\right)\ex^{\iu\omega t}+\left(\sigma_{n,n_1,n_2}^{m,-}(r),\epsilon_{n,n_1,n_2}^{m,-}(r)\right)\ex^{-\iu\omega t}\right]\bm{Y}_{n,n_1,n_2}^m(\theta,\varphi)
    \label{eq:stress-tensor-expanded-1}
\end{equation}
where summation is over $n\in[0,N]$,  $m\in[-n,n]$,  $n_1\in[n-1,n+1]$ and $n_2\in(n_1-1,n_1+1)$. Generally, rank-2 tensors have 9 independent components but as the stress and strain tensors are symmetric they only have 6 independent components. Because of this, it is convenient to use an alternative projection based on tensor harmonics defined by  Zerilli  \citep{Zerilli1970}. Following the notation of \cite{James1976};

\begin{equation}    \left(\bm\sigma'(r,\theta,\varphi,t),\bm\epsilon'(r,\theta,\varphi,t)\right)=\sum_{n,n_2,m,l}\left[\left(\sigma^{(l),+}_{n,n_2;m}(r),\epsilon^{(l),+}_{n,n_2;m}(r)\right)\ex^{\iu\omega t}+\left(\sigma^{(l),-}_{n,n_2;m}(r),\epsilon^{(l),-}_{n,n_2;m}(r)\right)\ex^{-\iu\omega t}\right]\bm{T}^{(l)}_{n,n_2;m}(\theta,\varphi).
\label{eq:stress-tensor-expanded-2}
\end{equation}
This projection has the advantage of separating the trace $l=0$, anti-symmetric $l=1$ and symmetric trace-free $l=2$ components. For a symmetric rank-2 tensor there are  $6$ non-zero components of the stress and strain tensor for each degree and order: the trace ($l=0$, $n_2=0$) and the symmetric spheroidal ($l=2$, $n_2=n$ and $n_2=n\pm2$) and toroidal ($l=2$, $n_2=n\pm1$) components. The Zerilli and rank-2 spherical harmonics tensor spherical harmonics are linearly related (see Appendix \ref{ap:rank-2-tensor}). 

To apply the boundary conditions, the radial components of the stress tensor are required.  The radial component of the stress tensor can be written in the rank-1 spherical harmonics basis or alternatively in the radial-poloidal-toroidal basis often used in literature, 
\begin{equation}    \bm{e}_r\cdot\bm{\sigma'}=\sum_{n,m}\left[R_n^{m,+}(r)\bm{R}_n^m(\theta,\varphi)+S_n^m(r)\bm{S}_n^{m,+}(\theta,\varphi)+T_n^{m,+}(r)\bm T_n^{m}(\theta,\varphi)\right]\ex^{\iu\omega t}+c.c. 
    \label{eq:stress-tensor-surface-decomposition-1}
\end{equation}
The spheroidal, poloidal and toroidal components can be written in terms of the $\sigma^{(l)}_{n,n_2;m}$ components of the stress tensor (Eq.~(\ref{eq:stress-radial-expanded})). 

The previous tensors must be real, which implies that the $+$ and $-$ components are related. Once the solution to the $+$ component is obtained the solution to the $-$ component follows immediately from the complex-conjugate properties of tensor spherical harmonics (see Appendix \ref{sec:ortogonality-properties})

\subsection{Spectral Form of the Governing Equations}
\label{sec:spectral-form-equations-motion}
Using the spectral expansions defined in Section \ref{subsec:spectral-expansion} and the properties of tensor spherical harmonics \citep{James1976}, we can transform the mass and momentum conservation equations and Poisson equation (Eq.~\ref{eq:equations-of-motion}) as well as the constitutive law (Eq.~\ref{eq:rheology}) to the spectral domain. 

The components of the momentum equations (Eq~.(\ref{eq:equations-of-motion}b)) in the $\bm{Y}^m_{n,n-1}$, $\bm{Y}^m_{n,n}$ and $\bm{Y}^m_{n,n+1}$ basis are given by
\begin{equation}
\begin{split}
    \sum_{n_2,l}(-1)^{n+n_2}\Lambda(n_1,l)\begin{Bmatrix}
1 & l & 1\\
n & n_1 & n_2
\end{Bmatrix}G(n_1,n_2)\partial_{n_2}^{n_1}\sigma_{n,n_2;m}^{(l)}-&\rho_0G(n,n_1)\sum_{n_{1\alpha}}G(n,n_{1\alpha})\partial_{n}^{n_{1\alpha}} (gu_{n,n_{1\alpha}}^m)-\\
g\rho_0G(n,n_1)\chi_n^m
-\rho_0G(n,n_1)\partial_n^{n_1}\phi_n^m=0
\end{split}
\label{eq:projected-momentum-equation}
\end{equation}
with $n_1\in(n-1,n+1)$. $G(n,n_1)$, $\partial_{n_1}^{n_2}$ , $\Lambda(n_1,l)$ are operators defined in Eqs.~(\ref{eq:derivative-operators}) and (\ref{eq:lambda})\added; and {the terms between curly brackets are Wigner's 6-j symbols}.

The Poisson equation (Eq.(\ref{eq:equations-of-motion}c)) is 
\begin{equation}
   D_n\phi_n^m=-4\pi G\rho\chi_n^m-4\pi G\partial_r\rho_0\sum_{n_{1\alpha}}G(n,n_{1\alpha})u^m_{n,n_{1\alpha}}
   \label{eq:projected-Poisson}
\end{equation}
where $\partial_r$ stands for the derivative with respect to the radial distance $r$, and $D_n$ is an operator defined in Eq.~(\ref{eq:D_operator}). The right-hand side is $0$ if the density is uniform within the layer and the solid is incompressible. The explicit forms of the momentum and the Poisson equation are given in Appendix \ref{ap:explicit-equations}.

To obtain the spectral form of the constitutive law (Eq.~(\ref{eq:rheology})), it is useful to rewrite it as

\begin{equation}
    \bm\sigma'=(3\lambda+2\mu)\bm T(\nabla\bm u)+2\mu\bm{S}(\nabla\bm u)=(3\lambda+2\mu)\bm T(\bm\epsilon')+2\mu\bm{S}(\bm\epsilon'),
\label{eq:rheology-2}
\end{equation}
$\bm{T}$ and $\bm{S}$ stand for the trace and trace-free symmetric parts of the tensor and we have used the definition of the strain tensor
\begin{equation}
    \bm{\epsilon'}=\frac{1}{2}\left(\nabla\bm{u}+(\nabla\bm{u})^\dag\right)=\bm T(\nabla\bm{u})+\bm S(\nabla\bm{u}).
    \label{eq:strain-trace-free-sym}
\end{equation}
An explicit expression for the strain tensor, which can be obtained using the properties listed in \S 3 of \cite{James1976}, is given in Appendix \ref{ap:explicit-equations}.

Plugging   Eq.~(\ref{eq:rheology-expanded}) into Eq.~(\ref{eq:rheology-2}) and using the spectral expansion of the strain tensor, we can obtain the stress tensor 

\begin{equation}
    \bm{\sigma'}=\left(\hat{A} _{0}+A_0\sum_{n_\beta,m_\beta}\hat {A}^{m_\beta}_{n_\beta}Y^{m_\beta}_{n_\beta}\right)    \left(\sum_{\substack{n_\alpha,n_{2\alpha},\\l_{\alpha},m_\alpha}}\epsilon_{n_{\alpha},n_{2\alpha};m_{\alpha}}^{(l_\alpha)}\bm{T}^{(l_{\alpha})}_{n_{\alpha},n_{2\alpha};m_\alpha}\right)
\label{eq:stress-tensor-expanded-3}
\end{equation}
with 
\begin{equation}
    \hat{A}=\begin{dcases}
    3\kappa,& \text{if } l=0\\
    2\hat\mu,& \text{if } l=2
\end{dcases}
\end{equation}
$\hat{A}_0$ and $A_0\hat{A}_n^m$ stand for the laterally-uniform and laterally-varying components of $A$, respectively. Finally, the components of the stress tensor $\sigma^{(l)}_{n,n_2;m}$ are obtained by projecting the stress tensor into a basis of Zerilli's tensors 
\begin{equation}
    \sigma^{(l)}_{n,n_2;m}=\frac{1}{4\pi}\int \bm{\sigma'}: \overline{{\bm{T}^{(l)}_{n,n_2;m}}}\text{d}\Omega
\end{equation}
where $:$ is the inner-product. Using Eq. (\ref{eq:stress-tensor-expanded-3}), we find
\begin{equation}
    \begin{split}
    \sigma^{(l)}_{n,n_2;m}=\hat{A}_0\epsilon_{n,n_2;m}^{(l)}+A_0
&\sum_{n_\beta,m_\beta}\hat{A}^{m_\beta}_{n_\beta}\sum_{\substack{n_\alpha,n_{2\alpha},\\l_\alpha,m_\alpha}}\epsilon_{n_\alpha,n_{2\alpha};m_\alpha}^{(l_{\alpha})}\left(\bm{T}^{(l_{\alpha})}_{n_{\alpha},n_{2\alpha};m_\alpha}:\overline{\bm{T}^{(l)}_{n,n_{2};m}}\cdot Y_{n_\beta}^{m_\beta}\right).
    \end{split}
\label{eq:projected-rheology}
\end{equation}
$\left(\bm{T}^{(l_{\alpha})}_{n_{\alpha},n_{2\alpha};m_\alpha}:\overline{\bm{T}^{(l)}_{n,n_{2};m}}\cdot Y_{n_\beta}^{m_\beta}\right)$ are the coupling coefficients, which indicate the coupling of mode ($n_\alpha,m_\alpha$) and $(n,m)$ due to rheology variations of degree and order ($n_\beta,m_\beta$). The coupling coefficients with non-zero value determine which modes are coupled. Their explicit form and properties are given in Appendix \ref{ap:coupling-coefficients}. 

Eqs.~(\ref{eq:projected-momentum-equation}), (\ref{eq:projected-Poisson}), (\ref{eq:strain-trace-free-sym}) and (\ref{eq:projected-rheology}) form a closed set of partial differential equations that can be solved under well-posed boundary conditions.

\subsection{Boundary Conditions}
\label{sec:boundary-condition}
To obtain the tidal response, the previous set of equations must be solved under appropriate boundary conditions. We impose boundary conditions at the surface $R$ and at the core-mantle boundary $R_2$. The derivation of the boundary conditions can be found in \cite{Sabadini2016}. At the surface, $R$,  the stress vanishes
\begin{equation}
    R_n^m=0\quad S_n^m=0,\quad T_{n}^m=0,
    \label{eq:bc-surf-1}
\end{equation}
and the potential stress, $Q$,is given by 
\begin{equation}
    Q_n^m=\partial_r\phi_n^m+\frac{n+1}{R}\phi_n^m+4\pi G\rho_3U_n^m=\frac{2n_T+1}{R}\delta_{n_T}^n\delta_{m_T}^m.
    \label{eq:bc-surf-2}
\end{equation}
with $n_T$ and $m_T$ the degree and order of the considered tidal forcing. 

 At the boundary between the liquid layer and the outermost solid layer ($R_2$), there is zero tangential and toroidal stresses and the radial stress is given by the difference between the radial displacement and the geoid
\begin{equation}
    S_n^m=0, \quad T_n^m=0;\quad R_n^m=\rho_2 g\left(U_n^m+\frac{\phi^m_n}{g}\right)
\end{equation}
The gradient of the gravitational potential is
\begin{equation}
    \partial_r\phi_n^m=\frac{n}{r}\phi_n^m+\frac{4\pi G}{g\rho_2}\left(\rho_3-\rho_2\right)\left(\rho_2\phi_n^m-R_n^m\right).
\end{equation}

Degree $0$ and $1$ require a special treatment \citep{Farrell1972,Qin2014}. For degree $0$  the displacement and  radial stress vector only have radial components. Because of compressibility, the body can experience non-zero radial displacements. However, as the mass of the body does not change, the perturbing potential should be $0$, $\phi_0=0$. The degree $1$ solution automatically satisfies $ R_1^m+2S_1^m+\frac{gQ_1^m}{4\pi G}=0$, which implies that two of the boundary conditions are a linear combination of each other. This is because the degree $1$ solution includes a translation of the center of mass that does not introduce stress. As we are working in the center of mass reference center we constrain this translation to be 0, $\phi_1^m=0$. Moreover, $T_1^m(R_2)=0$ is automatically fulfilled if $T_1^m(R)=0$. This is because the toroidal mode contains a net-rotation that does not introduce stress. Instead of using $T_n^m(R_2)=0$, we impose a $0$ toroidal displacement at $R_2$ without introducing extra stress $W_1^m(R_2)=0$, as done in \cite{Qin2014}.

\subsection{Love Numbers of an Aspherical Body}
\label{sec:Love-numbers-aspherical-body}
The tidal response of a body can be expressed in terms of Love numbers, a set of dimensionless proportionality constants that relate the tidal force and the body's response. For a spherically-symmetric body, a tidal forcing of a given degree and order results in a response of  the same degree and order. Moreover, due to  the spherical-symmetry, the Love numbers are independent of the order of the forcing. Because of this, there exist a set of frequency-dependent Love numbers ( radial $h_n$, poloidal $l_n$, and gravitational $k_n$) per degree that can be evaluated at any radial point. If there are lateral-heterogenities, a forcing of a given wavelength might excite a mode with different wavelength. Because of this the definition of Love numbers must be extended. More generally, we can write the tidal response as 
\begin{equation}
    (U,V,W,\phi)=\sum_{n_\alpha,m_\alpha}\sum_{n_\beta,m_\beta}\left(-\frac{h_{n_\alpha,m_\alpha}^{n_\beta,m_\beta}}{g_0},-\frac{l_{n_\alpha,m_\alpha}^{n_\beta,m_\beta}}{g_0},-\frac{t_{n_\alpha,m_\alpha}^{n_\beta,m_\beta}}{g_0},\delta_{n_\alpha}^{n_\beta}\delta_{m_\alpha}^{m_\beta}+k_{n_\alpha,m_\alpha}^{n_\beta,m_\beta}\right){{\phi^{(T)}}^{m_\alpha,+}_{n_\alpha}}Y_{n_\beta}^{m_\beta}\ex^{\iu\omega t}+c.c.
    \label{eq:Love-aspherical}
\end{equation}
$\alpha$ and $\beta$ indicate the tidal response at degree and order $n_\beta,m_\beta$ due to a forcing at degree and order $n_\alpha,m_\alpha$. $t$ are the toroidal  Love number, which are $0$ for a spherically-symmetric body. The only non-zero numbers Love numbers for a spherically-symmetric body are $h_{n_\alpha,m_\alpha}^{n_\alpha,m_\alpha}$, $l_{n_\alpha,m_\alpha}^{n_\alpha,m_\alpha}$, $k_{n_\alpha,m_\alpha}^{n_\alpha,m_\alpha}$, and as they are independent of order, can be simply written as  $h_n$, $l_n$, and $k_n$. 


\subsection{Energy Dissipation}
\label{sec:energy-dissipation}
If the body is anelastic, its deformation is not adiabatic. The volumetric energy dissipation is given by
\begin{equation}
    \dot{e}(r,\theta,\varphi)=\frac{1}{T}\int_t^{t+T} \bm{\sigma'}(r,\theta,\varphi,t):\bm{\dot\epsilon'}(r,\theta,\varphi,t)\text{d}t.
\label{eq:tidal-dissipation-0}
\end{equation}

The previous integral can be computed either in the temporal-spatial domain using $\bm\epsilon(t,r,\theta,\varphi)$  and $\bm\sigma(t,r,\theta,\varphi)$ or alternatively in the spectral domain using their spectral expansions. Plugging Eq.~(\ref{eq:stress-tensor-expanded-2}) into Eq.~(\ref{eq:tidal-dissipation-0}) and performing the time integral, we find

\begin{equation}
\begin{split}
    \dot{e}=\omega\iu\sum\limits_{\substack{n_\alpha,n_{2\alpha},l_\alpha \\ n_\beta,n_{2\beta},l_\beta\\m_\alpha,m_\beta}}\bm{T}_{n_\alpha,n_{2\alpha};m_\alpha}^{(l_\alpha)}:\bm{T}_{n_\beta,n_{2\beta};m_\beta}^{(l_\beta)}
   \left(\sigma_{n_\alpha,n_{2\alpha};m_\alpha}^{(l_\alpha),-}\epsilon_{n_\beta,n_{2\beta};m_\beta}^{(l_\beta),+}-\sigma_{n_\alpha,n_{2\alpha};m_\alpha}^{(l_\alpha),+}\epsilon_{n_\beta,n_{2\beta};m_\beta}^{(l_\beta),-}\right)
\end{split}
\label{eq:energy-r}
\end{equation}

For a given radius, we can project the previous expression into spherical harmonics to obtain the spectra of energy dissipation, i.e., 
\begin{equation}
    \dot{e}(r,\theta,\varphi)=\sum_{n,m}e_n^m(r)Y_n^m(\theta,\varphi)
    \label{eq:energy-spectra-ex}
\end{equation}
with 
\begin{equation}
    \dot{e}_{n_\nu}^{m_\nu}(r)=\omega\iu\sum\limits_{\substack{n_\alpha,n_{2\alpha},l_\alpha \\ n_\beta,n_{2\beta},l_\beta\\m_\alpha,m_\beta}}
   \left(\sigma_{n_\alpha,n_{2\alpha};m_\alpha}^{(l_\alpha),-}\epsilon_{n_\beta,n_{2\beta};m_\beta}^{(l_\beta),+}-\sigma_{n_\alpha,n_{2\alpha};m_\alpha}^{(l_\alpha),+}\epsilon_{n_\beta,n_{2\beta};m_\beta}^{(l_\beta),-}\right)\left(\bm{T}^{(l_\alpha)}_{n_{\alpha},n_{2\alpha};m_\alpha}:{\bm{T}^{(l_\beta)}_{n_\beta,n_{2\beta};m_\beta}}\cdot\overline{Y_{n_\nu}^{m_\nu}}\right)
   \label{eq:energy-surface-pattern}
\end{equation}
$\left(\bm{T}^{(l_{\alpha})}_{n_{\alpha},n_{2\alpha};m_\alpha}:{\bm{T}^{(l_\beta)}_{n_\beta,n_{2\beta};m_\beta}}\overline{Y_{n_\nu}^{m_\nu}}\right)$ is an integral defined in Appendix \ref{ap:coupling-coefficients}.
The average energy dissipation for radius $r$ is given by $\dot{e}_0^0(r)$ and can be obtained using the orthogonality and complex conjugate properties of tensor spherical harmonics (Appendix \ref{sec:ortogonality-properties}) 
\begin{equation}
\begin{split}
    \dot{e}_0^0(r)=
   2\omega\sum\limits_{\substack{n,n_{2},l,m}}  \left[\Im(\sigma_{n,n_{2},l}^{m,+})\Re(\epsilon_{n,n_{2},l}^{m,+})-\Re(\sigma_{n,n_{2},l}^{m,+})\Im(\epsilon_{n,n_{2},l}^{m,+})\right]
\end{split}
\label{eq:total-energy}
\end{equation}
The total energy dissipation can be found by radially integrating $\dot e_0^0$,
\begin{equation}
    \dot{E}=\int_V\dot{e}\textrm{d}V=\int_S\int_{r_1}^{r_2}\sum_{n,m}\dot{e}_n^m(r)Y_n^m r^2\text{d}\Omega\text{d}r=4\pi\int_{r_1}^{r_2}\dot{e}_0^0(r)r^2\text{d}r.
    \label{eq:energy-method-1}
\end{equation}

Alternatively, the total tidal dissipation can be obtained from the work done by the tidal force \citep{love1927treatise}
\begin{equation}
    \dot{E}=\frac{1}{T}\int_V\int_T \rho\frac{\partial\bm{u}}{\partial t}\cdot\nabla\phi^T\text{d}t\text{d}V,
    \label{eq:energy-method-2}
\end{equation}
which can be transformed into a surface integral \citep[e.g.,][]{Zschau1978,PEALE1978,Platzman1984}
\begin{equation}
    \dot{E}=\frac{\omega}{8\pi^2GR}\sum_n(2n+1)\int_T\int_S\phi^{T}_n\frac{\partial\phi'_n}{\partial t}\text{d}t\text{d}S
\end{equation}
where $\phi'$ is the perturbing potential (i.e, $\phi=\phi'+\phi^{T}$), and the subscript $n$ indicates that only the component of degree $n$ of the potential is considered. Using the spectral expansion of the tidal potential (Eq.~(\ref{eq:generic-forcing})), the definition of the Love numbers (Eq.~\ref{eq:Love-aspherical}) and the orthogonality of spherical harmonics, we find: 
\begin{equation}
    \dot{E}=-2\frac{\omega R}{G}\sum_{\substack{n_1,m_1,\\n_2,m_2}}(2n+1){\phi^{(T)}}^{m_1,+}_{n_1}{\phi^{(T)}}^{m_2,+}_{n_2}\Im(k_{n_2,m_2}^{n_1,m_1}). 
    \label{eq:energy-method-2-2}
\end{equation}
If we consider a spherically-symmetric body in an eccentric orbit, the previous expression reduces to the classic expression \citep[e.g.,][]{peale1979}:
\begin{equation}
    \dot{E}=-\frac{21}{2}\frac{(\omega R)^5 e^2}{G}\Im(k_{2}).
\end{equation}

\section{Numerical Approach}
\label{sec:numerical-method}
The system of differential equations of Section \ref{sec:spectral-form-equations-motion}  is written in terms of the $3$ components of the displacement vector ($u^m_{n,n-1},u^m_{n,n},u^m_{n,n+1}$), the $6$ components of the strain $(\epsilon^{(0)}_{n,n;m},\epsilon^{(2)}_{n,n-2;m},\epsilon^{(2)}_{n,n-1;m},\epsilon^{(2)}_{n,n;m},\epsilon^{(2)}_{n,n+1;m},\epsilon^{(2)}_{n,n+2;m})$ and stress $(\sigma^{(0)}_{n,n;m},\sigma^{(2)}_{n,n-2;m},\sigma^{(2)}_{n,n-1;m},\sigma^{(2)}_{n,n;m},\sigma^{(2)}_{n,n+1;m},\sigma^{(2)}_{n,n+2;m})$ tensors, and the gravitational potential ($\phi_n^m$) for each degree and order. However, it is more convenient to rewrite the equations in terms of the variables traditionally employed to solve the viscoelastic deformation of a self-gravitating body \citep[e.g.,][]{Love1911,Farrell1972,Sabadini2016}: the radial, poloidal and toroidal components of the displacement vector and of the radial component of the stress tensor ($U_n^m,V_n^m,W_n^m,R_n^m,S_n^m,T_n^m$) and the gravitational potential and its gradient ($\phi,\partial_r \phi$): 
\begin{equation}
\begin{split}
    \bm{y}=&(U_0, R_0, \phi_0, \partial_r\phi_0, \ldots U^{m}_{n}, V_n^m, R^m_n, S_n^m, \phi_n^m,\partial_r\phi_n^m, W_n^m, T_n^m \ldots U^{\infty}_{\infty}, V_\infty^\infty, R^\infty_\infty, S_\infty^\infty, \phi_\infty^\infty,\partial_r\phi_\infty^\infty, W_\infty^\infty, T_\infty^\infty)\\
    &=(y_{0,1}, y_{0,3}, y_{0,5}, y_{0,6}, \ldots y^{m}_{n,1}, y^{m}_{n,2}, y^{m}_{n,3}, y^{m}_{n,4}, y^{m}_{n,5}, y^{m}_{n,6}, y^{m}_{n,7} y^{m}_{n,8} \ldots y^{\infty}_{\infty,1}, y_{\infty,2}^\infty, y^\infty_{\infty,3}, y_{\infty,4}^\infty, y_{\infty,5}^\infty,y_{\infty,6}^\infty, y_{\infty,7}^\infty, y_{\infty,8}^\infty),
\end{split}
\end{equation}
where we have also introduced the $y$ radial functions traditionally used in the viscoealstic-gravitational problem \citep[e.g.,][]{Sabadini2016}. We note that the problem is formulated in terms of the gradient of the gravitational potential $y_6=\partial_r\phi$ rather than the potential stress $Q$. The equations can be then be cast in the form (Appendix \ref{ap:explicit-equations})
\begin{equation}
    \partial_r \bm{y}(r)=\bm{\mathrm{D}}\bm{y}(r),
\end{equation}
where $\bm{\mathrm{D}}$ is a linear operator that depends on the interior properties and the radial distance $r$. 

Given a forcing of degree $n_T$ and order $m_T$ not all the modes ($n_T$, $m_T$) are excited. If there are no lateral rheology variations, equations of different degree and order are decoupled.. Additionally, if we only consider a zonal forcing $m_T=0$ and no longitudinal rheology variations (i.e., $\kappa_n^m=\mu_n^m=\eta_n^m=0$ for $m\neq 0$), spheroidal ($U_{n}^m,V_n^m,R_n^m,S_n^m,\phi_n^m,\partial_r\phi_n^m$) and toroidal ($W_{n}^m,T_n^m$) modes are decoupled. More generally, the modes $(n,m)$ involved in the tidal response depend on the forcing spectra $(n_{T},m_{T})$  and that of the lateral variations $(n_\textrm{LV},m_\textrm{LV})$ via the coupling terms. Given a mode ($n_\alpha,m_\alpha$) and lateral variations of the form $(n_\textrm{LV},m_\textrm{LV})$ the coupled modes $(n_\alpha,m_\alpha)\bigotimes(n_\textrm{LV},m_\textrm{LV})\Rightarrow(n,m)$ are given by the non-zero coupling coefficients. The coupled modes can be obtained recursively by using the selection rules listed in Appendix \ref{ap:coupling-coefficients}. Starting from the tidal force ($n_T,m_T$) the modes involved in the solution can be obtained by recursively applying the selection rules: first order modes,  $(n_{T},m_{T})\bigotimes(n_\textrm{LV},m_\textrm{LV})\Rightarrow(n_1,m_1)$, second order modes $(n_1,m_1)\bigotimes(n_\textrm{LV},m_\textrm{LV})\Rightarrow(n_2,m_2)$, and modes of order $p$, $(n_{p-1},m_{p-1})\bigotimes(n_\textrm{LV},m_\textrm{LV})\Rightarrow(n_p,m_p)$. 

As opposed to a spherically symmetric body, for which there is only one mode involved in the tidal response per forcing harmonic $(n_T,m_T)$, an infinite set of modes are excited by the tidal force when lateral variations are considered. This makes it impossible to obtain an exact solution to the problem using the spectral method. An approximate solution is obtained by setting a maximum cut-off order $N_p$. The reduced system of differential equations is given by
\begin{equation}
    \partial_r \bm{y}_a(r)=\bm{\mathrm{D}}_a\bm{y}_a(r).
    \label{eq:diff-eq}
\end{equation}
with $\bm{y}_a$ containing only the considered modes and $\bm{\mathrm{D}}_a$ their corresponding dynamics. $\bm{y}_a$ is a $8N_{\mathrm{modes}}$ vector or $8(N_{\mathrm{modes}}-1)+4$ vector if the degree $0$ response is excited. To obtain the tidal response, Eq.~(\ref{eq:diff-eq}) is integrated numerically from the boundary between the liquid layer and the outermost solid layer ($R_2$) to the surface using a Runge-Kutta scheme. The solution is given by
\begin{equation}
    \bm{y}_a=\sum_{n,m,k}C_{n,k}^m \bm{y}_{n,k}^m.
    \label{eq:solution-y}
\end{equation}
$C_{n,k}^m$ are a set of $8N_{\mathrm{modes}}$, $8(N_{\mathrm{modes}}-1)+4$ if the degree $0$ is excited, integration constants; and $\bm{y}_{n,k}^m$ are the corresponding set of solution vectors with ${y}_{n',k'}^{m'}(R_2)=\delta_{n}^{n'}\delta_{m}^{m'}\delta_{k}^{k'}$.

The integration constants are obtained by applying the boundary conditions listed in Section \ref{sec:boundary-condition} and the solution follows from Eq.~(\ref{eq:solution-y}). Once the radial functions $y$ are obtained, the components of the displacement vector, and strain and stress tensors in tensor spherical harmonic basis are obtained using Eqs.~(\ref{eq:displacement-vector-rad-tan}), (\ref{eq:strain-tensor-explicit}) and (\ref{eq:stress-tensor-explicit}). The solution can be transformed from the spectral domain to the spatial and time domain using Eqs.~(\ref{eq:rank-0-expansion}), (\ref{eq:rank-1-expansion}) and (\ref{eq:stress-tensor-expanded-2}) and the definition of the rank 0 (Eq.~(\ref{eq:rank-0-def})), rank 1 (Eq.~(\ref{eq:rank-1-def})) and rank 2 (Eq.~(\ref{eq:rank-2-def})) tensor spherical harmonics. 

\begin{sidewaysfigure}[ht] 
\plotone{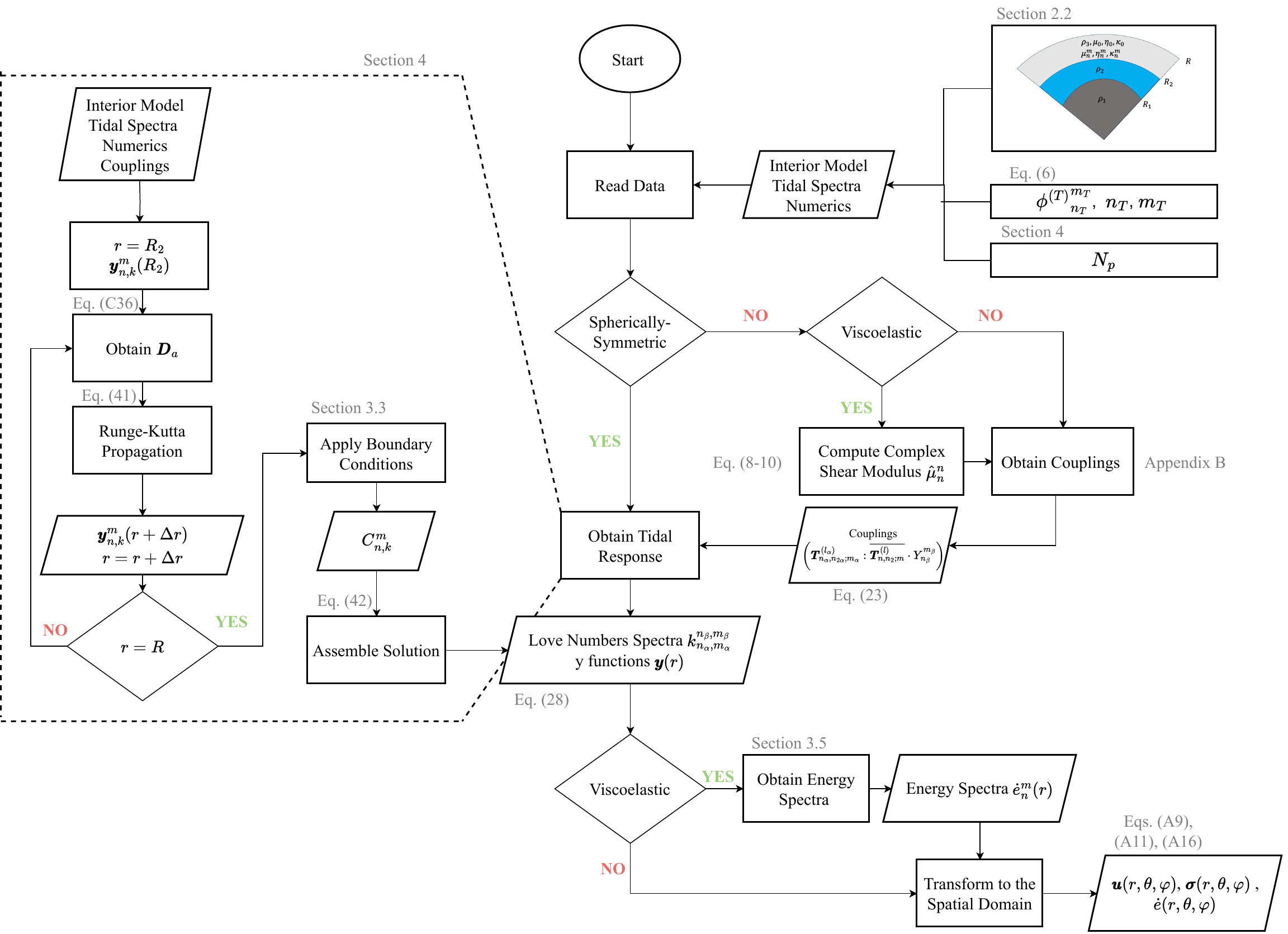}
\caption{ Flowchart of the methodology employed to obtain the tidal response} 
\label{fig:flow-chart}
\end{sidewaysfigure}

Figure \ref{fig:flow-chart} provides an overview of the method described above, which is implemented in the \textit{LOV3D} software repository.

\section{ Model Benchmark and Comparison with Previous Methods}

\label{sec:model-benchmark}
We compare the results obtained with the spectral method presented here with those obtained using the spectral-perturbation method of \cite{Qin2014,Qin2016} and the FEM code of \cite{Berne2023a,Berne2023b}. 

The spectral method of \cite{Qin2014,Qin2016} relies on perturbation theory. As opposed to the method presented above, the solution is computed recursively. The tidal response of the spherically-symmetric body is used to obtain  the first-order modes response, which is then employed to compute second-order modes, etc. For each mode, only the radial functions corresponding to that mode are considered unknown. The coupling terms (second term in the RHS of Eq.~(\ref{eq:projected-rheology})) are considered known and follow from a lower order mode, effectively acting as a forcing. This way, the equations corresponding to each of the modes participating in the tidal response are decoupled. This means that for a mode of a given perturbation order, the effect of higher order modes is ignored. 

The spectral-perturbation method of \cite{Qin2014,Qin2016} provides an approximated solution whose accuracy is expected to decrease as the amplitude of lateral variations increases. To test the accuracy of the perturbation method, we consider an elastic Io-like body (Table \ref{tab:non-dimensional-numbers}, $\tau\rightarrow\infty$) with zonal ---$(n_\textrm{LV},m_\textrm{LV})=(1,0)$, $(n_\textrm{LV},m_\textrm{LV})=(2,0)$--- and sectoral ---$(n_\textrm{LV},m_\textrm{LV})=(1,\pm1)$--- lateral variations under a zonal tidal forcing --- $(n_\textrm{T},m_\textrm{T})=(2,0)$--- and compute the tidal response. We note that similar results are obtained for different interior parameters and forcing. For each of the considered lateral heterogeneities, we obtain the gravitational Love numbers using the two methods and compute the difference between them. The tidal response can be written as the tidal response of a spherically symmetric body, given by $k_{2}^u$, plus an additional response arising from lateral variations $\Delta k_{2,0}^{n,m}$. 

\begin{figure}[ht!]
\plotone{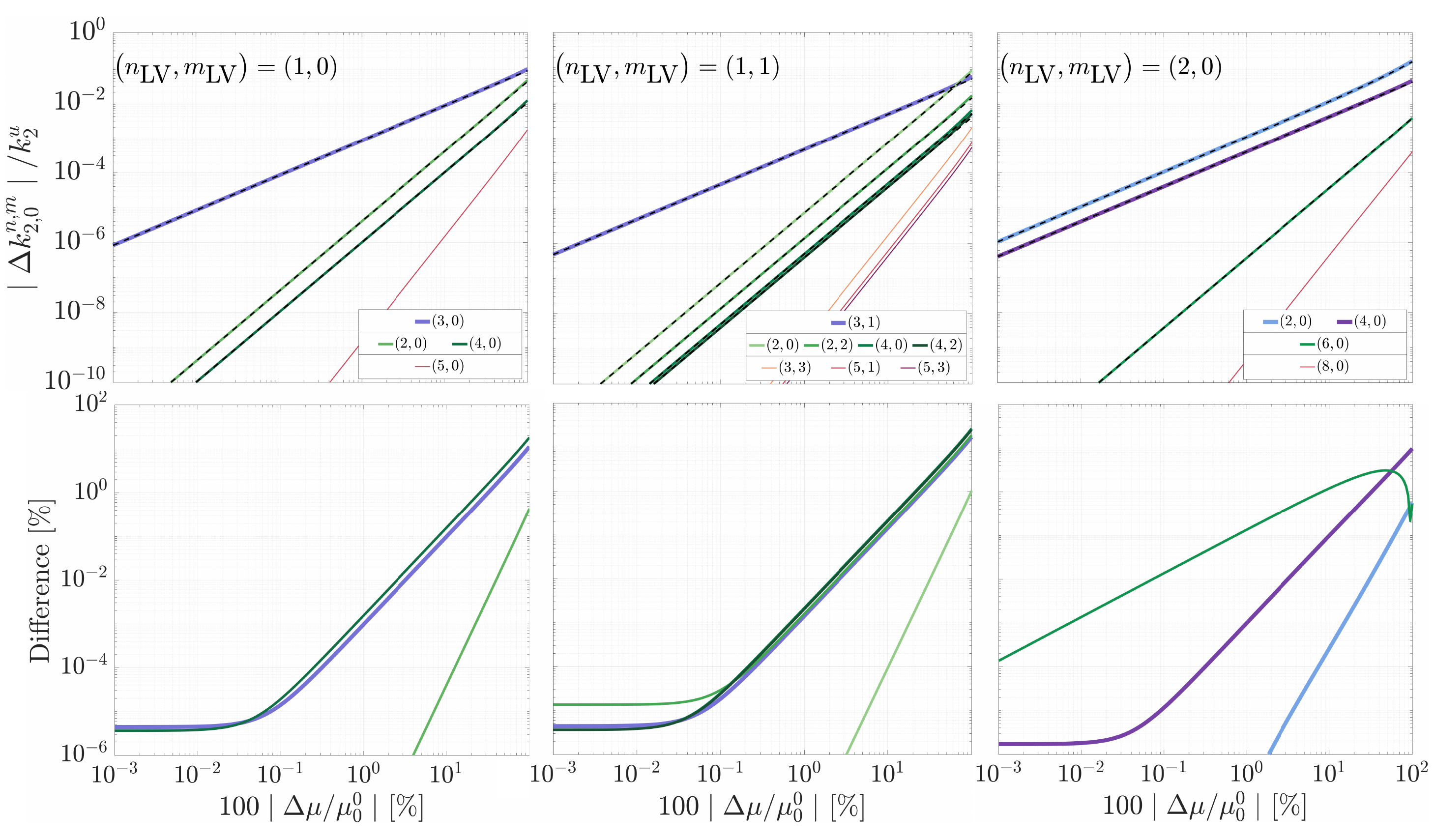}
\caption{Gravitational Love numbers for different types of lateral shear modulus variations (upper panels); and difference with the Love numbers obtained with the perturbation approach  of \cite{Qin2014} (lower panels). An elastic Io (Table \ref{tab:non-dimensional-numbers}) with shear modulus variations of different wavelengths $(n_\mathrm{LV},m_\mathrm{LV})$ is assumed. The amplitude of lateral variations is given in terms of the peak-to-peak variation of the shear modulus with respect to its mean. The solution is cut-off at perturbation order 3. The line-width indicates the order of the mode, with thicker lines indicating lower order modes. In the upper panels the dashed lines corresponds to the solution obtained with the perturbation method of \cite{Qin2014}. For $(n_\textrm{LV},m_\textrm{LV})=(1,1)$ only the $+\lvert m\rvert$ modes are shown, $-\lvert m\rvert$ modes of amplitude $k_{2,0}^{n,- \lvert m \rvert}=(-1)^{\lvert m \rvert}k_{2,0}^{n,\lvert m \rvert}$ are also excited. The difference between the two methods is computed as $100\left\lvert\left( k_{2,0}^{n,m}/k_2^u\right)_R-\left( k_{2,0}^{n,m}/k_2^u\right)_Q\right\rvert/\left\lvert k_{2,0}^{n,m}/k_2^u\right\rvert_R$, with $R$ and $Q$ standing for the method presented here and the method of \cite{Qin2014}, respectively.} 
\label{fig:benchmark-Qin}
\end{figure}

Figure \ref{fig:benchmark-Qin} shows the tidal Love number spectra for the three sets of lateral heterogeneities; first, second and third order modes are indicated. The additional tidal response increases from being $0.01\%$ of the tidal response of the uniform body for peak-to-peak shear modulus variations of $\sim 0.1\%$ the mean shear modulus to about $10\%$ for variations of the same order of magnitude as the mean shear modulus. The additional tidal response is clearly dominated by first-order modes, which for small lateral variations are orders of magnitude higher than second-order modes. However, the difference between first and second order modes decreases as the amplitude of lateral variations increases. 

The results obtained with the spectral and the spectral-perturbation method show good agreement. Differences between the results obtained with the two models are as small as $\sim 10^{-5}\%$ for lateral variations of less than $\sim 10^{-3}\%$. The difference is likely caused by differences in how the equations are radially-integrated; for instance, the difference decreases as the number of radial points used in the Runge-Kutta integration increases. This difference increases as the amplitude of lateral heterogeneities increases. However, it remains small even when substantial lateral heterogenites are considered. For $10\%$ peak-to-peak variations of the shear modulus the difference remains below $1\%$, discrepancies of $10\%$ or more are only attained when peak-to-peak variations of the shear modulus have the same order of magnitude as the mean shear modulus. The remarkable agreement is due to the fast decrease of mode amplitude with increasing mode order, as shown in Figure \ref{fig:benchmark-Qin}a,c and d. This means that the effect of modes of higher order modes on modes of lower order remains small even with high-amplitude lateral heterogeneities. 

We also consider the case of an icy moon. We obtain the tidal response for the reference Enceladus model (Table \ref{tab:non-dimensional-numbers}) for various types of lateral variations using both spectral methods and the FEM code. For the FEM approach, we adapt the methodology outlined in \cite{Berne2023a} and \cite{Berne2023b} to develop a FEM capable of simulating tidal deformation on laterally heterogeneous ocean worlds. FEMs solve the equation of motion for quasi-static problems by formulating 3D displacements (i.e., in response to applied forces and boundary conditions) as a series of linear shape functions across a mesh domain. For this work, we discretize mesh domains using tetrahedra with a maximum edge length of 1\si{km} and consider between $18\cdot10^{6}$ and $33\cdot10^{6}$ elements for each simulation. To simulate tidal loading, we upload mesh geometries to a modified version of the geodynamic software package Pylith \citep{Aagaard2007PyLith:Deformation} which can self-consistently consider forces arising from external tidal potentials, self-gravity, and radial displacements at boundaries between internal density layers (for additional details, see Supplementary S1.1 of \cite{Berne2023a}). To incorporate lateral variations in elastic properties, we sample analytic basis functions (i.e., spherical harmonics) for a given heterogeneity at mesh node locations. Following simulations, we expand displacements into spherical harmonics and compute Love numbers for comparison to semi-analytic solutions presented in this work.

\begin{figure}[ht!]
\plotone{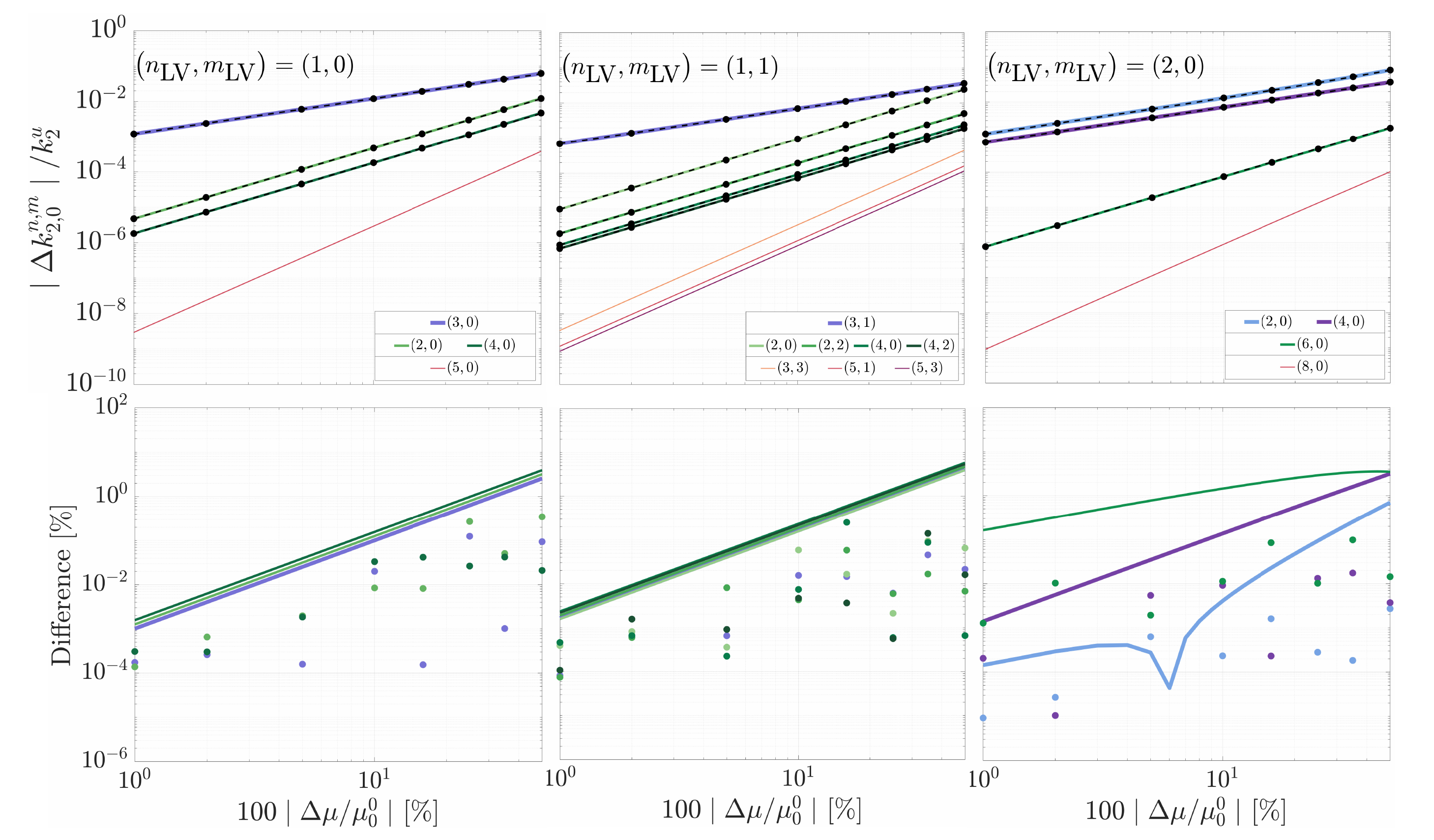}
\caption{Same as Figure \ref{fig:benchmark-Qin} but for the Enceladus model (Table \ref{tab:non-dimensional-numbers}). Differences are obtained as $100\left\lvert\left( \Delta k_{2,0}^{n,m}/k_2^u\right)_R-\left( \Delta k_{2,0}^{n,m}/k_2^u\right)_X\right\rvert/\left\lvert\Delta k_{2,0}^{n,m}/k_2^u\right\rvert_R$, where $X$ can either be the results obtained using the FEM model or the spectral perturbation method. Solutions obtained using the FEM of \cite{Berne2023a} and the perturbation method of \cite{Qin2014} are indicated with filled circles and dashed lines, respectively.}
\label{fig:benchmark-Qin-FEM}
\end{figure}

Figure \ref{fig:benchmark-Qin-FEM} compares the results obtained with the three methods (i.e., spectral method, spectral-perturbation method and FEM). With few exceptions, the results obtained with the FEM  and the spectral method show the best agreement. As explained above, the difference between the spectral and spectral perturbation method grows with the amplitude of lateral variations reaching values as high as $5\%$ for peak-to-peak shear modulus variations of $50\%$. In contrast, the difference between results obtained with the spectral and the FEM remains below $1\%$ and is often one order of magnitude smaller than for the former. We ascribe small discrepancies between results with the spectral method and the FEM code to errors associated with resolution. We find that the agreement between both methods improves when the FEM mesh size decreases from 5 to 1 \si{km}.

To benchmark the viscoelastic component, we compute the Love numbers of a spherically-symmetric uniform body obtained using analytical expressions \cite[e.g.,][Appendix B]{Matsuyama2018} and \textit{LOV3D} and find excellent agreement (relative differences of $<10^{-6}\%$). For a viscoelastic body with lateral variations we check that energy dissipation is computed self-consistently by comparing the total energy dissipation obtained using the two approaches outline in Section \ref{sec:energy-dissipation} (i.e., Eqs.~(\ref{eq:energy-method-1}) and (\ref{eq:energy-method-2-2})), finding good agreement (see Appendix \ref{ap:energy-dissipation-test}).

\section{The Tides of a Laterally-Heterogeneous Elastic Body}
\label{sec:elastic-body}

\begin{figure}
\gridline{\fig{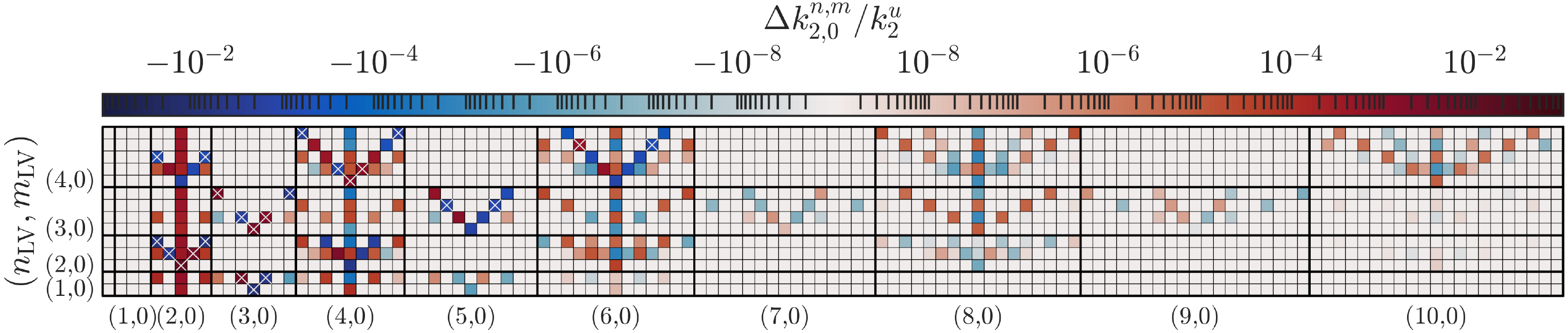}{0.98\textwidth}{}
          }
          \vspace{-1.2cm} 
\gridline{\fig{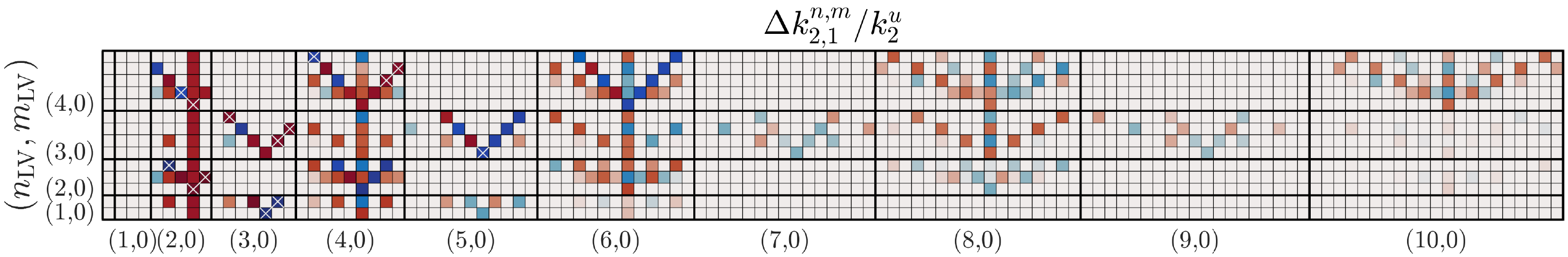}{0.98\textwidth}{}
          }
          \vspace{-1.2cm} 
\gridline{\fig{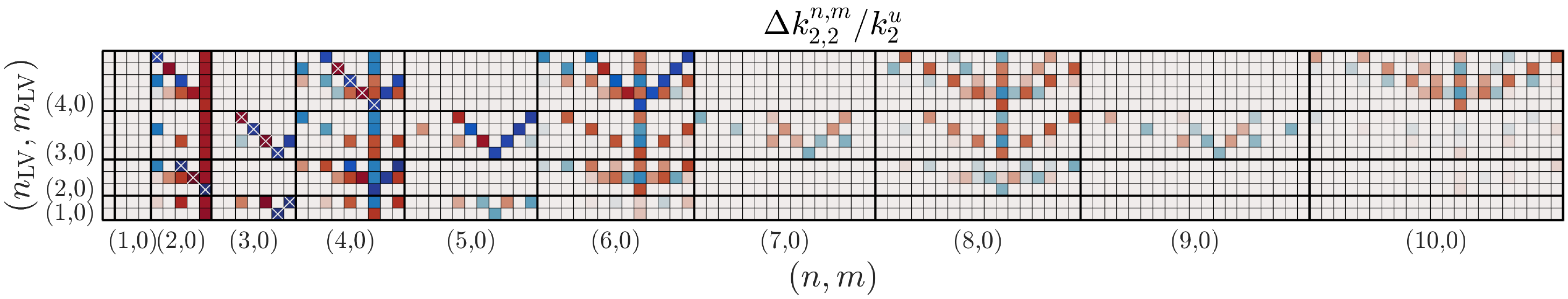}{0.98\textwidth}{}
          }
          \vspace{-1cm} 
\caption{Love number spectra for a degree two order zero (upper panel), one (center panel) and two (lower panel) tidal forcing for lateral variations of different degrees and orders ($n_\mathrm{LV},m_\mathrm{LV}$). An elastic Io is assumed (Table \ref{tab:non-dimensional-numbers}), and the solution is cut-off at perturbation order 4. Color intensity indicates the strength of the mode, and the terms with highest amplitude are indicated with a cross. For all cases, we assume peak-to-peak shear modulus variations of $10\%$ of the mean shear modulus. }
\label{fig:Love-spectra}
\end{figure}

Figure \ref{fig:benchmark-Qin} evidences that different spatial patterns of lateral variations results in a distinct tidal response. This is further illustrated in Figure \ref{fig:Love-spectra}, which shows  the degree $2$ gravitational Love spectra for \textit{monochromatic} (single total and zonal wave-number) shear modulus variations. Each type of lateral variations leads to a unique Love number spectra. Hence, if the full tidal response was measured the inverse problem could be solved and the spatial pattern and amplitude of lateral variations inferred, this technique is known as \textit{tidal tomography}.

The use of tidal tomography presents several challenges. The amplitude of the additional tidal response arising from lateral variations is small, making it challenging to measure. Because of this, we can expect that only the terms with highest amplitudes will be measured. Moreover, lateral variations in different regions and of different properties (e.g., elastic properties, layer thickness) can result in a similar tidal spectra. Finally, lateral variations are likely not  monochromatic but feature various wavelengths. This makes the solution of the inverse problem non-unique. As an example, both degree $1$ and $3$ lateral variations produce a degree $3$ tidal signal (see Figure \ref{fig:Love-spectra}). If no other terms of the spectra are measured, a degree $3$ response would be indicative of a hemispherical dichotomy but could not be used to distinguish between degree $3$ and $1$ lateral variations. Distinguishing between the two would require measuring the degree $5$ tidal response, which is more prominent for degree $3$ lateral variations. Lateral variations also alter the tidal response at the degree and order of the forcing, most prominently for degree $2$ variations. This makes it challenging to distinguish the effects of lateral heterogeneity from the tidal response and might even result in errors when estimating the mean properties of the body. In such cases, a way to detect lateral variations is by comparing the $k_2$ Love numbers at different orders, which are only equal if the body is spherically symmetric. The tidal potential of a moon in an eccentric orbit has both order $0$ and $2$ components, making this approach attractive. 

\begin{figure}[ht!]
\plotone{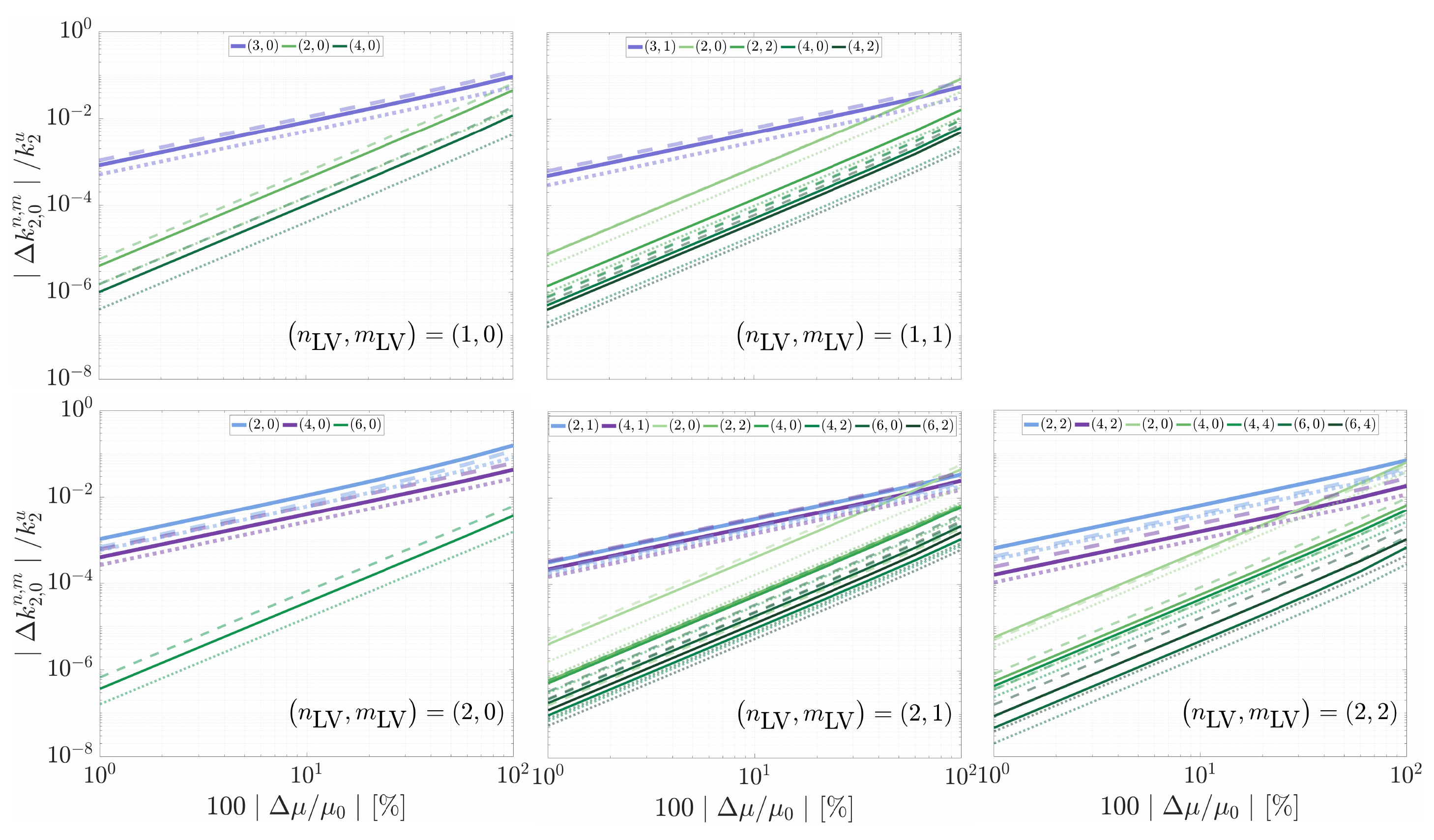}
\caption{Gravitational Love numbers, $k_{2,0}^{n,m}$, for lateral shear modulus variations of different wavelengths $(n_\mathrm{LV},m_\mathrm{LV})$ for Io (solid lines), the Moon (dashed lines) and Mercury (dotted lines).} 
\label{fig:rocky}
\end{figure}

So far, tidal tomography has just been used for Earth \citep{Lau2017}, for which there is high quality geodetic data. However, future space missions might make it possible to use tidal tomography for other Solar System bodies. Figure \ref{fig:rocky} shows the gravitational Love number spectra for Io, the Moon and Mercury (Table \ref{tab:non-dimensional-numbers}). In all cases, the additional tidal response ($\Delta k$) is normalized by the tidal response of the spherically-symmetric body ($k_2^u$), listed in Table \ref{tab:non-dimensional-numbers}. 

The Gravity Recovery and Interior Laboratory (GRAIL) mission provided very accurate lunar gravity data, which can be used to constrain the amplitude of lateral heterogenities in the Lunar interior \citep{Qin2014}. We caution that a comprehensive inversion can only be done if an expression for the Love numbers that account for the effect of lateral variations (i.e., Eq~(\ref{eq:Love-aspherical})) rather than the classic expression often used \cite[e.g.,][Eq.~(17)]{Konopliv2013} is used when obtaining the gravity solution from raw GRAIL data, and the model parameter space is thoroughly sampled. However, the results shown in Figure \ref{fig:rocky} combined with the available GRAIL gravity field give a qualitative impression of the amplitude of lateral variations. For instance, \cite{Williams2014} showed that the measured degree $3$ response of the Moon is consistent with a spherically-symmetric model, this suggests that there are not high amplitude odd-degree lateral variations, as otherwise one would measure a high degree $3$ response. Additionally, the degree $2$ orders $0$ and $2$ Love numbers of the Moon differ by $\sim 1\%$, but are within the uncertainty of the measurements \citep{Konopliv2013,Lemoine2013}. From Figure \ref{fig:rocky}, this suggests that degree two variations are less than $\sim 10\%$. 

Of the three bodies, Mercury is the one for which the additional tidal response is the smallest relative to the tidal response of a spherically-symmetric. Nevertheless, as Mercury is the body with the highest $k_2^u$ , lateral variations lead to the highest gravity signal in absolute terms. The \textit{MESSENGER} mission measured Mercury's $k_2$ with an accuracy of approximately $5\%$, insufficient to observe lateral variations. In contrast, \textit{BepiColombo}, scheduled to start its science operations in $2026$, is expected to improve the accuracy to $\sim0.1\%$ \citep{Genova2021}. The accuracy at which the non-diagonal Love numbers can be obtained (e.g., $k_{n_\alpha,m_\alpha}^{n_\beta,m_\beta}$, with $n_\alpha\neq n_\beta$, $m_\alpha\neq m_\beta$) can only be determined via a simulation of \textit{BepiColombo}'s gravity science experiment. Assuming the accuracy inferred for $k_2$ extends to the other Love numbers, variations of the shear modulus or of the thickness of Mercury's rocky envelope as small as $2\%$ might be detected. As demonstrated by Figure \ref{fig:rocky}, the detection threshold depends on the spatial pattern of lateral variations.

\begin{figure}[ht!]
\plotone{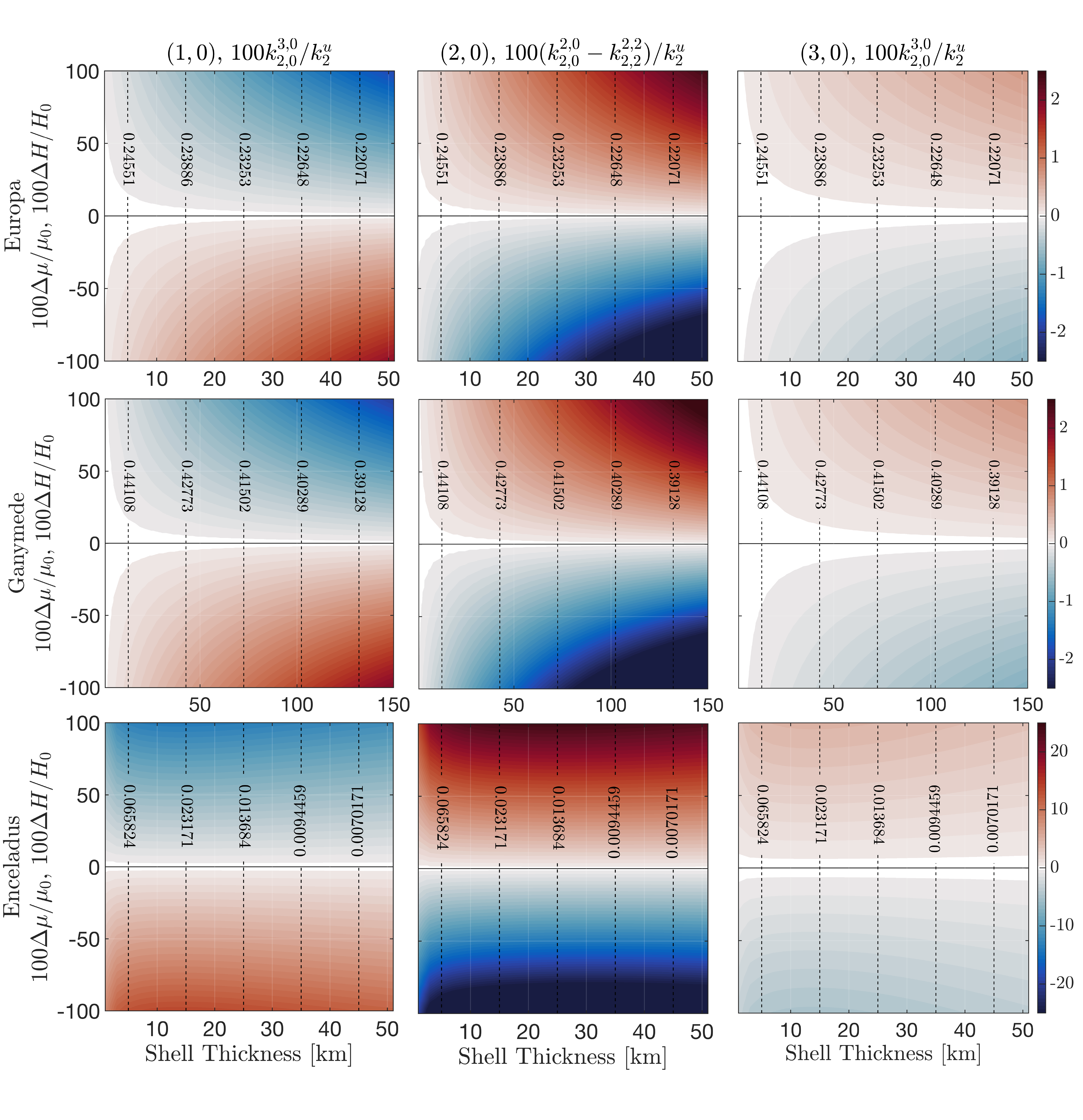}
\caption{Effect of lateral heterogeneities of different wavelengths for Europa (first row), Ganymede (second row) and Enceladus (third row). The highest amplitude component of the tidal response arising from zonal lateral variations of degrees $1$ (first column), $2$ (second column) and $3$ (third column) is depicted. For reference, the degree two Love number of a spherically symmetric body, $k_2^u$, is indicated (dashed black lines). } 
\label{fig:non-dimensional}
\end{figure}

We also consider the role of lateral heterogeneities in icy moons. As noted in Section \ref{sec:non-dimensional-numbers}, shell thickness variations can be approximately mapped to shear modulus variations provided the shell is thin compared to the moon's radius. This is the case of icy moons ---for Europa and Ganymede the shell to radius ratio is $\sim0.02$ and $0.05$, respectively; for Enceladus it is higher $\sim0.1$, making the approximation less accurate. Figure \ref{fig:non-dimensional} shows the additional tidal response for zonal variations of the shear modulus for Europa, Ganymede and Enceladus. The relative additional tidal response resulting from lateral variations ($\Delta k_{n_T,m_T}^{n,m}/k_{n_T}^u$) is approximately one order of magnitude higher for Enceladus than for the two Jovian moons. For Enceladus, lateral variations of $50\%$ cause an additional tidal response $\sim10\%$ the amplitude of the main tidal response (see also \cite{Berne2023a} and \cite{Behounkova2017}), while for the latter the additional tidal response is $\sim1\%$. Nevertheless, as $k_{2}$ of Enceladus is more than one order of magnitude smaller (Table \ref{tab:non-dimensional-numbers}), $\Delta k_{n_T,m_T}^{n,m}$ is similar for the three icy moons. 

The possibility to observe lateral variations depends on their amplitude and the accuracy to which the Love number spectra can be obtained. For Enceladus, gravity and shape data indicates that shell thickness varies from $29$ \si{km} at the equator to $7$ \si{km} at the south pole \citep{Beuthe2016,Cadek2016,HEMINGWAY2019}. In contrast, shell thickness variations for Europa are not expected to exceed $7$ \si{km} \citep{NIMMO2007}. Ganymede's shell thickness variations are not constrained, yet ocean circulation models predict shell thickness variations to be smaller for big icy moons \citep{Kang_2022}. This makes Enceladus a prime candidate for the use of tidal tomography in the future. With the JUpiter Icy Moons Explorer (\textit{JUICE}) on its way to the Jovian system, it is interesting to consider if it can pick up the signal arising from lateral variations of shell properties. \textit{JUICE} will measure Ganymede's $k_2$ with an expected accuracy of $10^{-4}$ \citep{CAPPUCCIO2020}, which is less than $0.1\%$ of Ganymede's expected $k_2$. This accuracy is sufficient to detect differences in $k_{2,0}^{2,0}$ and $k_{2,2}^{2,2}$ caused by lateral variations of ice shell properties, making tidal tomography a promising tool for constraining lateral variations in Ganymede's shell.

\section{The Tides of a Laterally-Heterogeneous Maxwell Body}
\label{sec:viscoelastic}
In the previous section, we considered the tidal response of an elastic body with lateral variations. We will now consider how lateral variations affect the tidal response of a viscoelastic body. If a body is not perfectly elastic, part of the tidal energy is converted into heat. Even for a spherically-symmetric body, tidal heating is not homogeneously distributed within the body's interior. Here we explore the effect that lateral variations of the viscosity has on tidal heating patterns.   

We employ our reference Io model (see Table \ref{tab:non-dimensional-numbers}) and consider various patterns of viscosity variations. The complex shear modulus spectra is obtained using Eqs.~(\ref{eq:complex-shear-1})-(\ref{eq:complex-shear-2}). Even in the case of monochromatic viscosity variations the complex shear modulus spectra is not monochromatic but contains infinite terms of varying amplitude. To use the spectral method, the complex shear modulus spectra needs to be cut-off. We consider terms up to $2$ orders of magnitude smaller than the leading non-degree-0 term of the complex shear modulus. The heating pattern depends on the tidal potential spectra. We focus on the tides experienced by a synchronous body in an eccentric orbit with zero obliquity --- for which we give the tidal potential in Appendix \ref{sec:tidal-potential-synch-sat}. We obtain the tidal response for each component of the tidal potential and compute the energy spectra as described in Section \ref{sec:energy-dissipation}. 

\begin{sidewaysfigure}[ht] 
\plotone{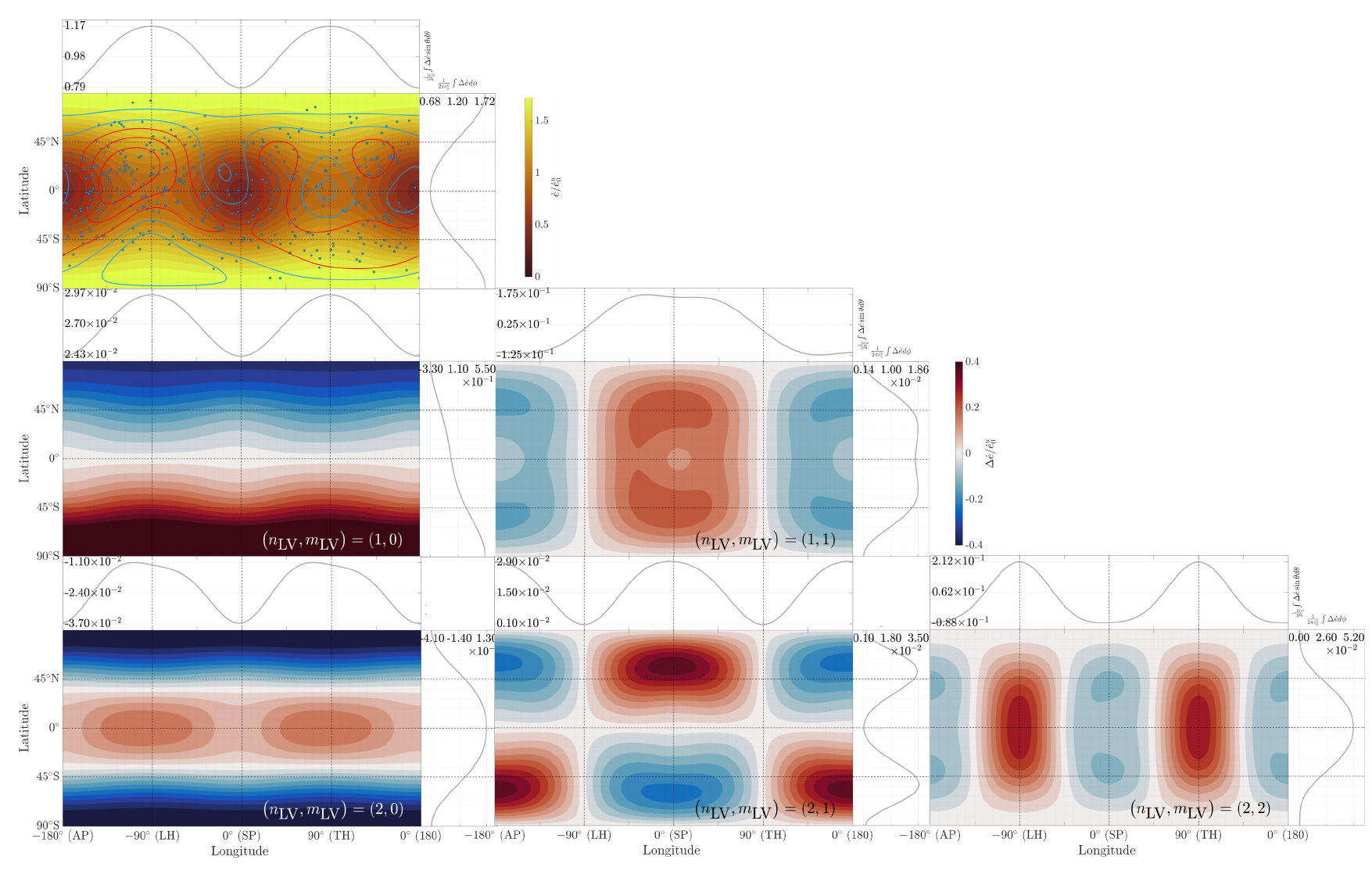}
\caption{Pattern of tidal heating for a spherically-symmetric body (upper row) and for bodies with different types of lateral viscosity variations. For the spherically-symmetric body the total heating pattern is shown while for the rest the difference between their corresponding heating pattern and that of a spherically-symmetric body is shown. All plots are normalized by the average tidal heating of the spherically-symmetric body. The reference Io model is employed (see Table \ref{tab:non-dimensional-numbers}) with peak to peak viscosity  variations of $50\%$. The sub-planet (SP) and anti-planet (AP) points as well as the centers of the trailing and leading hemispheres (TH and LH) are indicated. The longitudinally- and longitudinally-averaged tidal heating are also shown. In the first panel, Io's volcanoes are indicated (blue dots), with contours corresponding to the density of Io's volcanoes (red and blue contours indicating higher and lower than the global average)  \citep{Steinke2020_Pattern}. } 
\label{fig:tidal-heating-patterns}
\end{sidewaysfigure}

\begin{figure}[ht!]
\plotone{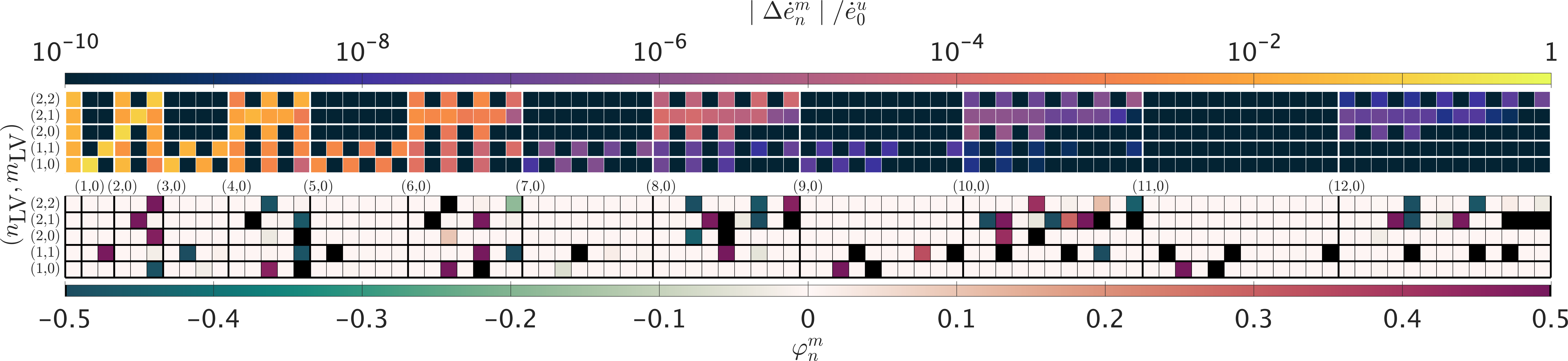}
\caption{Tidal heating spectra (Eq.~(\ref{eq:energy-spectra-ex})) corresponding to the cases of Figure \ref{fig:tidal-heating-patterns}. The energy spectra is given in terms of amplitude (upper panel) and longitudinal eastward shift in fraction of longitudinal wavelength with respect to $\Re(Y_n^m)$ (lower panel).} 
\label{fig:tidal-heating-spectra}
\end{figure}

Figures \ref{fig:tidal-heating-patterns} and \ref{fig:tidal-heating-spectra} show the spatial distribution of tidal heating for various types of monochromatic lateral viscosity variations and its corresponding spectra (Eq.~\ref{eq:energy-spectra-ex}), respectively. The tidal heating pattern characteristic of a spherically-symmetric body consists of terms of degree and order $(0,0)$, $(2,0)$, $(2,\pm 2)$, $(4,0)$, $(4,\pm2)$ and $(4,\pm4)$ \citep[e.g.,][]{beuthe2013}. For the three layer model considered here (solid core, liquid core and rocky envelope), tidal heating is maximum at the poles; along the equator, tidal heating exhibits minima at the sub-planet ($0^\circ$) and anti-planet ($180^\circ$) points and maxima at the center of the trailing and leading hemispheres ($90^\circ$ and $-90^\circ$, respectively). Viscosity lateral variations introduce additional terms. The difference between the tidal heating pattern of a uniform body and a body with lateral variations is dominated by the same pattern as the considered viscosity variations. For example, a polar dichotomy of degree $1$ and order $0$ in viscosity translates into the same dichotomy in tidal heating (see second panel in  Figure \ref{fig:tidal-heating-patterns}). This also follows from the tidal heating spectra, where the most prominent terms are those with same wavelength as the considered viscosity variations (Figure \ref{fig:tidal-heating-spectra}). Apart from this dominant term, other terms with smaller amplitude also arise. 

Some terms of the tidal heating spectra are not symmetric with respect to the $0^\circ$ meridian. This is most evident by looking at the tidal heating map corresponding to the degree $2$ order $0$ lateral variations, where we observe a westward shift in the longitude at which the minimum heat flux is attained compared to the spherically-symmetric case. The appearance of a trailing-leading hemispheres asymmetry is remarkable since the considered patterns of lateral variations do not contain such an asymmetry. A similar phenomenon was observed by \cite{Steinke_Chapter_3}, who used FEM to obtain tidal heating patterns for a multi-layered, laterally-heterogeneous Io. The asymmetry arises because the tidal potential is not symmetric with respect to the east-west direction. As shown in Appendix \ref{sec:tidal-potential-synch-sat}, the tidal potential can be broken into a degree $2$ order $0$ standing wave and degree $2$ order $2$ westward and eastward propagating waves of different amplitudes. The amplitude of the eastward component is greater than that of the westward component, breaking the symmetry of the problem. No asymmetry is observed if a single standing-wave is considered. 

Tidal heating patterns can help constraining the interior properties of rocky and icy worlds \citep[e.g.,][]{Breuer2022}. The distribution of volcanoes in Io has been used as a proxy of tidal heating. In Io, the concentration of volcanoes is greater towards mid- to low-latitudes and bimodal, exhibiting two maxima $30^\circ-60^\circ$ eastward of the sub-Jovian and anti-Jovian points (see Figure \ref{fig:tidal-heating-patterns}). Moreover, the distribution of volcanoes also contains a statistically-significant degree-6 component \citep{HAMILTON2013,KIRCHOFF201122,Steinke2020_Pattern}. The link between tidal heating and volcanic patterns is not simple: convection and melt transport affects how the two patterns relate \citep{STEINKE2020}, and the inferred volcanic patterns are affected by biases in observations. Recent observations by \textit{JUNO} have discovered new volcanoes in previously poorly-covered polar regions and even hinted that the north pole could have a higher concentration of hot-spots \citep{Zambon2023,Davies2024}.  

The tidal heating pattern characteristic of a 2-layer (core+rocky envelope) spherically-symmetric Io does not match the observed distribution of volcanoes. As illustrated in Figure \ref{fig:tidal-heating-patterns}, the surface heat flux peaks at the poles, does not exhibit a $30^\circ-60^\circ$ eastward shift with respect to the tidal axis, does not have degree-6 terms nor a polar dichotomy. The heating pattern is affected by both radial and lateral variations of internal properties. The presence of a low-viscosity asthenosphere can account for the concentration of volcanoes at mid- to low-latitudes. However, for spherically-symmetric cannot reproduce the observed eastward shift of volcanic activity with respect to the sub-Jovian point. \cite{Tyler2015} showed that tidal heating in a magma ocean can cause this shift. The results discussed here, as well as the FEM model of \cite{STEINKE2020}, indicate that lateral viscosity variations can introduce a longitudinal shift. The magnitude and direction of the longitudinal shift depends on the particular form of lateral and radial variations of internal properties. A comprehensive exploration of the joint effect of radial and lateral variations is left for future work. As evidenced in Figure \ref{fig:tidal-heating-spectra}, lateral variations of internal properties can also introduce a degree $6$ heat flux pattern and cause a polar-dichotomy (terms with odd $m$). 

Interior properties and heating patterns are tightly related. Viscosity and shear modulus depend on temperature and melt fraction. Lateral variations of tidal heating affect both quantities \citep{STEINKE2020} and hence the heating pattern. This feedback can have important implications for the interior evolution of tidally-active bodies. As shown in Figure \ref{fig:tidal-heating-spectra}, even a body with a uniform interior exhibits non-uniform tidal heating. Such a pattern will alter material properties and feedback into the heating pattern, driving interior evolution. Alternatively, primordial lateral heterogeneities might be amplified by the feedback in a similar way. The extent to which this happens depends on how heat is transported within the body. Tackling this complex problem requires coupling a tidal and a thermal model. The lower computational cost of spectral method compared to FEM methods makes them an attractive tool to approach the problem.

\section{Summary and Outlook}
\label{sec:discussion-and-conclusions}
In this work, we presented a spectral method to compute the tidal response of bodies with lateral heterogeneities and applied it to elastic and viscoelastic bodies. Below, we summarize the key points: 
\begin{itemize}
    \item The new spectral method shows good agreement with results obtained with the spectral-perturbation approach of \cite{Qin2014} and the FEM model of \cite{Berne2023a,Berne2023b}. Differences between the results obtained with the new method and the perturbation method grow as the amplitude of lateral variations increase. Predictions obtained using the perturbation method differ by less than $1\%$  for shear modulus variations of  $\lesssim 10\%$ and only reach $\sim 10\%$ for peak-to-peak variations of the same order of magnitude as the mean shear modulus. This makes the perturbation method a powerful tool to compute the tidal response of bodies unless such variations have high amplitude. Differences between the FEM approach and the presented spectral method remain small even for high amplitude lateral variations. Spectral methods are more computationally-efficient: a model that takes $\sim 10$ days to run using the FEM code in a two-core computer is solved in less than a minute using the spectral method presented here; in terms of memory, the grids employed in the FEM runs presented here required $\sim 1$ \si{GB} of memory, while the coupling matrices employed in the spectral model required $\sim 1$ \si{MB}. This makes spectral methods more suitable to tackle the inverse problem. However, unlike FEMs, the spectral code presented here cannot include faults and cracks, non-linear rheology, and layer thickness variations can only be treated by mapping them to shear modulus variations \citep{Berne2023a}.

    \item Each set of lateral variations results in distinct Love number spectra. Thus, measurements of the complete tidal response of a body can be used to constrain lateral variations. Nevertheless, the solution of the inverse problem is non-unique as it is unlikely that the complete Love number spectra can be measured and lateral variations at different depths might result in similar spectra (Section \ref{sec:elastic-body}). Other measurements, such as static gravity and topography, might help to solve this degeneracy. 

    \item For the Moon, Io and Mercury, shear modulus variations of the same order of magnitude as the mean shear modulus can cause an additional tidal response as high as $\sim 1-10\%$ the main tidal response. \textit{BepiColombo} should observe the fingerprint of lateral variations provided they are higher than approximately $2\%$ the mean shear modulus (Section \ref{sec:elastic-body}).

    \item In icy worlds, lateral variations of ice shell thickness modify the tidal response. Due to  the expected amplitude of shell-thickness variations, Enceladus is a prime candidate to use tidal tomography. For Europa and Ganymede, shell-thickness variations can lead to an additional tidal response in the order of $0.1-1\%$ the main tidal response depending on their amplitude. The accuracy of \textit{JUICE} makes it possible to detect this signal (Section \ref{sec:elastic-body}).

    \item Lateral variations modify the distribution of tidal heating for viscoelastic bodies. The additional tidal heating pattern due to lateral variations of the viscosity is dominated by the pattern of such variations. Lateral viscosity variations can cause a trailing-leading hemisphere asymmetry in tidal heating. This could explain the eastward shift of volcanic activity with respect to the sub-Jovian and anti-Jovian points observed in Io (Section \ref{sec:viscoelastic}).  

    \item  The dependence of interior properties on temperature and melt fraction, which in turn depend on the distribution of tidal heating, gives rise to a complex feedback that can drive interior evolution. The computational efficiency of the spectral method makes it a good candidate to study this feedback (Section \ref{sec:viscoelastic}). 

    \item For this work, we considered a simplified interior structure consisting of a solid non-deformable core, overlaid by an hydrostatic liquid and a solid envelope with radially-uniform properties (Section \ref{sec:non-dimensional-numbers}). These assumptions can be relaxed ---i.e., the deformation of the inner core can be considered, dynamic liquid tides included and radial variations of interior properties introduced.
    
\end{itemize}

\vspace{1cm}
M.R. and I.M. were supported by the National Aeronautics and Space Administration (NASA) under grant No. 80NSSC20K0570 issued through the NASA Solar System Workings program. A.B was supported by the Future Investigators in NASA Earth and Space Science and Technology (FINESST) Program (80NSSC22K1318). The authors thank Allard Veenstra for his contributions to the \textit{LOV3D} software and for providing feedback on the original manuscript. The \textit{LOV3D} software can be downloaded  and run at \url{https://github.com/mroviranavarro/LOV3D_open}.

\software{
LOV3D \citep{LOV3D},
Wigner 3j-6j-9j \citep{Wigner}, cmocean \citep{cmocean}, M\_Map \citep{M_Map}, export\_fig \cite{export_fig}, Tidal-Response \cite{Qin2016_code}
}

\appendix
\section{Tensor Spherical Harmonics}
\label{ap:tensor-spherical-harmonics}
Tensor spherical harmonics are a generalization of spherical harmonics to tensors of rank $>0$. For problems with spherical geometry, expanding the tensors in tensor spherical harmonics allows to employ the properties of tensor spherical harmonics to eliminate longitude and latitude from the equations of motion.  A rank $k$ tensor $\bm{a}$ can be written as \citep{James1976}
\begin{equation}
    \bm{a}=\sum_{\substack{n,m,\\n_1,..,n_k}}a^m_{n,n_1,..n_k}\bm{Y}^m_{n,n_1,..n_k}.
\end{equation}

Below, we define tensor spherical harmonics and introduce some of their main properties. We also give explicit expressions of tensor spherical harmonics of ranks $0$ , $1$ and 2; explain how to transform between them and other basis often used in literature and in this manuscript; and list operators used in the text. An exhaustive list of properties can be found in \cite{James1976}.

\subsection{Definitions}
\label{sec:ortogonality-properties}
Tensor spherical harmonics of degree $n$, order $m$ of rank $k$ are recursively defined as  \citep{James1976}
\begin{equation}
    \bm{Y}^m_{n,n_1..n_k}=(-1)^{n-m}\Lambda(n)\sum_{m_1,\mu}\begin{pmatrix}
n & n_1 & 1\\
m & -m_1 & -\mu
\end{pmatrix}\bm{Y}_{n_1,..n_k}^{m_1}\bm{e}_\mu
\label{eq:tensor-spherical-harmonic-def}
\end{equation}
with the $2\times3$ array being Wigner 3-j coefficient and $\Lambda(a,b,c..)$ is given by
\begin{equation}
    \Lambda(a,b,c,...)=((2a+1)(2b+1)(2c+1)...)^{1/2}.
    \label{eq:lambda}
\end{equation}
$\bm{e}_\mu$ are defined in terms of the unit vectors of Cartesian coordinates ($\bm{1}_x$,$\bm{1}_y$,$\bm{1}_z$) as
\begin{equation}
   \bm{e}_0=\bm{1}_z\quad \bm{e}_{-1}=(\bm{1}_x-\iu\bm{1}_y)/\sqrt{2}\quad\bm{e}_{1}=-(\bm{1}_x+\iu\bm{1}_y)/\sqrt{2}
\end{equation}

The Wigner-3j symbols are only not zero when $n,n_1..n_k$ satisfy the triangular identity, which means that a rank $k$ tensor spherical harmonic of degree and order $n$ and $m$ have $3^k$ tensor spherical harmonics. 

For convenience, we will sometimes use $\bm{Y}_{n_\alpha(k)}$ to refer to $\bm{Y}^{m_\alpha}_{n_\alpha,n_{1\alpha},..,n_{k\alpha}}$

Tensor spherical harmonics form a orthogonal basis. Integrated over the unit sphere, we have that 
\begin{equation}
    \frac{1}{4\pi}\int_\Omega \bm{Y}_{n_\alpha(k)}\cdot\bm{Y}_{n_\beta(k)}{}^*\textrm{d}\Omega=\delta_{n_\alpha(k)}^{n_\beta(k)}
    \label{eq:orthogonal}
\end{equation}
where ${}^*$ indicates the complex-conjugate-transpose. $\delta_{n_\alpha(k)}^{n_\beta(k)}$ is defined as $\delta_{m_\alpha}^{m_\beta}\delta_{n_\alpha}^{n_\beta}\delta_{n_{1\alpha}}^{n_{1\beta}}\hdots\delta_{n_{k\alpha}}^{n_{k\beta}}$, where $\delta_{a}^{b}$ is the Kronecker delta; and $\cdot$ is the generalized dot product of two tensors defined in \cite{James1976} as
\begin{equation}
    \bm{e}_{\mu_1}\hdots \bm{e}_{\mu_k}\cdot\overline{\bm{e}}_{\lambda_k}\hdots \overline{\bm{e}}_{\lambda_1}=\delta_{\mu_1}^{\lambda_1}\hdots\delta_{\mu_k}^{\lambda_k}.
\end{equation}

The complex conjugate of a tensor spherical harmonic of rank $k$ is given by 
\begin{equation}
 \overline{\bm{Y}^m_{n(k)}}=(-1)^{n+n_k+m+k}\bm{Y}^{-m}_{n(k)} 
\label{eq:complex-conjugate-tensor-spherical-harmonic}
\end{equation}
which implies that if a tensor is real, the $m$ and $-m$ components are related as

\begin{equation}
 \overline{a^m_{n(k)}}=(-1)^{n+n_k+m+k}a^{-m}_{n(k)} 
\label{eq:complex-conjugate-tensor-spherical-harmonic-2}
\end{equation}

\subsection{Rank 0 Tensors}
\label{ap:rank-0-tensor}
Rank $0$ tensor spherical harmonics are the classic spherical harmonics $Y^n_m$
\begin{equation}
    Y^n_m(\theta,\varphi)=\sqrt{\frac{(2n+1)(n-m)!}{(n+m)!}}P_{n}^m(\cos\theta)\ex^{\iu m \varphi}
    \label{eq:rank-0-def}
\end{equation}
where $P_{n}^m$ are associated-Legendre functions 
\begin{equation}
    P_{n}^m(x)=(-1)^{n+m}\frac{(1-x^2)^{m/2}}{2^n n!}\frac{\textrm{d}^{m+n}(1-x^2)^n}{\textrm{d}x^{m+n}}
\end{equation}

\subsection{Rank 1 Tensors}
\label{ap:rank-1-tensor-spherical-harmonics}
Rank-1 tensor, such as the displacement vector $\bm u$, are expanded using tensor spherical harmonics of rank $1$, which can be written in spherical coordinates as

\begin{subequations}
    \begin{equation}
       \sqrt{n(2n+1)}\bm{Y}^m_{n,n-1}=nY\bm{e}_r+E\bm{e}_\theta+G\bm{e}_\varphi, 
    \end{equation}
    
    \begin{equation}
        \sqrt{n(n+1)}\bm{Y}^m_{n,n}=\iu G\bm{e}_\theta-\iu E\bm{e}_\varphi, 
    \end{equation}
    
    \begin{equation}
        \sqrt{(n+1)(2n+1)}\bm{Y}^m_{n,n+1}=-(n+1)Y\bm{e}_r+E\bm{e}_\theta+G\bm{e}_\varphi,
    \end{equation}
    \label{eq:rank-1-def}
\end{subequations}
 with

\begin{equation}
    Y=Y_n^m,\quad\quad E=\frac{\partial Y}{\partial \theta}, \quad F=\frac{\partial^2 Y}{\partial \theta^2},\quad G=\csc\theta\frac{\partial Y}{\partial \varphi},\quad H=\frac{\partial G}{\partial \varphi};
    \label{eq:YEFGH-defs}
\end{equation}

Rank 1 tensor spherical harmonics are related to the scaloidal–poloidal–toroidal basis ($\bm R_n^m$,$\bm S_n^m$, $\bm T_n^m$) as

\begin{subequations}
\begin{equation}
    \bm{R}_n^m=Y\bm{e}_r=\sqrt{\frac{n}{2n+1}}\bm{Y}^m_{n,n-1}-\sqrt{\frac{n+1}{2n+1}}\bm{Y}^m_{n,n+1}
\end{equation}

\begin{equation}
    \bm{S}_n^m=E\bm{e}_\theta+G\bm{e}_\varphi=\sqrt{\frac{n}{2n+1}}(n+1)\bm{Y}^m_{n,n-1}+\sqrt{\frac{n+1}{2n+1}}n\bm{Y}^m_{n,n+1}
\end{equation}

\begin{equation}
    \bm{T}_n^m=G\bm{e}_\theta-E\bm{e}_\varphi=\nabla_{\theta,\varphi}Y_n^m\times {\bm{1}_r}=-\iu\sqrt{n(n+1)}\bm{Y}^m_{n,n}
\end{equation}
\end{subequations}

We can transform between the components of vector in the rank-1 tensor spherical harmonics basis, $(u^m_{n,n-1},u^m_{n,n},u^m_{n,n+1})$, and in the scaloidal-poloidal-toroidal basis, $(U_n^m,V_n^m,W_n^m)$ as 
\begin{subequations}
\begin{equation}
    U_{n}^m=\frac{1}{\sqrt{2n+1}}\left[\sqrt{n}u^n_{n,n-1}-\sqrt{n+1}u^n_{n,n+1}\right],
\end{equation}
\begin{equation}
    V_n^m=\frac{1}{\sqrt{2n+1}}\left[\frac{u^n_{n,n-1}}{\sqrt{n}}+\frac{ u^n_{n,n+1}}{\sqrt{n+1}}\right],
\end{equation}
\begin{equation}
    W_n^m=\frac{\iu}{\sqrt{n(n+1)}}u^m_{n,n}.
\end{equation}
\label{eq:displacement-vector-rad-tan}
\end{subequations}
, or in matrix form
\begin{equation}
\bm{U}_{n}^m=\matrixB{C}_{u\rightarrow U}\bm{u}_{n,n_1}^m
\end{equation}
where we have introduced $\bm u_{n,n_1}^m=(\hdots u^m_{n,n-1},u^m_{n,n},u^m_{n,n+1}\hdots)^\dag$ and $\bm U_{n}^m=(\hdots U_n^m, V_{n}^m, W_n^n\hdots)^\dag$

\subsection{Rank 2 Tensors}
\label{ap:rank-2-tensor}
A rank-2 tensor ($\bm{\mathrm{A}}$) can be expanded into tensor spherical harmonics of rank 2, $\bm{\mathrm{A}}=\sum_{n,m,n_1,n_2}A^m_{n,n_1,n_2}\bm{Y}^m_{n,n_1,n_2}$, which in spherical coordinates are given by

\begin{subequations}
\begin{equation}
    \begin{split}
    &\sqrt{n(n-1)(2n-1)(2n+1)}\bm{Y}^m_{n,n-1,n-2}=
    n(n-1)Y\bm{e}_r\bm{e}_r
    +(n-1)E\bm{e}_r\bm{e}_\theta
    +(n-1)G\bm{e}_r\bm{e}_\varphi
    +(n-1)E\bm{e}_\theta\bm{e}_r\\
    &+(nY+F)\bm{e}_\theta\bm{e}_\theta
    +H\bm{e}_\theta\bm{e}_\varphi
    +(n-1)G\bm{e}_\varphi\bm{e}_r
    +H\bm{e}_\varphi\bm{e}_\theta
    -(F+n^2Y)\bm{e}_\varphi\bm{e}_\varphi
    \end{split}
\end{equation}

\begin{equation}
    \begin{split}
    &\iu n\sqrt{(n-1)(2n+1)}\bm{Y}^m_{n,n-1,n-1}=
    (1-n)G\bm{e}_\theta\bm{e}_r
    -H\bm{e}_\theta\bm{e}_\theta
    +(F+n^2Y)\bm{e}_\theta\bm{e}_\varphi\\
    &+(n-1)E\bm{e}_\varphi\bm{e}_r
    +(F+nY)\bm{e}_\varphi\bm{e}_\theta
   +H\bm{e}_\varphi\bm{e}_\varphi
    \end{split}
\end{equation}

\begin{equation}
    \begin{split}
    &n\sqrt{4n^2-1}\bm{Y}^m_{n,n-1,n}=
   -n^2 Y\bm{e}_r\bm{e}_r
    -nE\bm{e}_r\bm{e}_\theta
    -nG\bm{e}_r\bm{e}_\varphi
    +(n-1)E\bm{e}_\theta\bm{e}_r
    +(nY+F)\bm{e}_\theta\bm{e}_\theta\\
    &+H\bm{e}_\theta\bm{e}_\varphi
    +(n-1)G\bm{e}_\varphi\bm{e}_r
    +H\bm{e}_\varphi\bm{e}_\theta
    -(F+n^2Y)\bm{e}_\varphi\bm{e}_\varphi
    \end{split}
\end{equation}

\begin{equation}
    \begin{split}
    &\iu n \sqrt{(2n+1)(n+1)}\bm{Y}^m_{n,n,n-1}=
    -nG\bm{e}_r\bm{e}_\theta
    +nE\bm{e}_r\bm{e}_\varphi
    +G\bm{e}_\theta\bm{e}_r\\
    &-H\bm{e}_\theta\bm{e}_\theta
    +F\bm{e}_\theta\bm{e}_\varphi
    -E\bm{e}_\varphi\bm{e}_r
    +(F+n(n+1)Y)\bm{e}_\varphi\bm{e}_\theta
    +H\bm{e}_\varphi\bm{e}_\varphi
    \end{split}
\end{equation}

\begin{equation}
    \begin{split}
    n(n+1)\bm{Y}^m_{n,n,n}=
    -E\bm{e}_\theta\bm{e}_r
    +(F+n(n+1))\bm{e}_\theta\bm{e}_\theta
    +H\bm{e}_\theta\bm{e}_\varphi
    -G\bm{e}_\varphi\bm{e}_r
    +H\bm{e}_\varphi\bm{e}_\theta
    -F\bm{e}_\varphi\bm{e}_\varphi
    \end{split}
\end{equation}

\begin{equation}
    \begin{split}
    &\iu(n+1)\sqrt{n(2n+1)}\bm{Y}^m_{n,n,n+1}=
    (n+1)G\bm{e}_r\bm{e}_\theta
    -(n+1)E\bm{e}_r\bm{e}_\varphi
    +G\bm{e}_\theta\bm{e}_r\\
    &-H\bm{e}_\theta\bm{e}_\theta
    +F\bm{e}_\theta\bm{e}_\varphi
    -E\bm{e}_\varphi\bm{e}_r
    +(F+n(n+1)Y)\bm{e}_\varphi\bm{e}_\theta
    +H\bm{e}_\varphi\bm{e}_\varphi
    \end{split}
\end{equation}

\begin{equation}
    \begin{split}
    &(n+1)\sqrt{(2n+3)(2n+1)}\bm{Y}^m_{n,n+1,n}=
    -(n+1)^2Y\bm{e}_r\bm{e}_r
    +(n+1)E\bm{e}_r\bm{e}_\theta
    +(n+1)G\bm{e}_r\bm{e}_\varphi
    -(n+2)E\bm{e}_\theta\bm{e}_r\\
    &+(F-(n+1)Y)\bm{e}_\theta\bm{e}_\theta
    +H\bm{e}_\theta\bm{e}_\varphi
    -(n+2)G\bm{e}_\varphi\bm{e}_r
    +H\bm{e}_\varphi\bm{e}_\theta
   -(F+(n+1)^2)\bm{e}_\varphi\bm{e}_\varphi
    \end{split}
\end{equation}

\begin{equation}
    \begin{split}
    &\iu(n+1)\sqrt{(n+2)(2n+1)}\bm{Y}^m_{n,n+1,n+1}=
    (n+2)G\bm{e}_\theta\bm{e}_r
    -H\bm{e}_\theta\bm{e}_\theta
    +(F+(n+1)^2)Y\bm{e}_\theta\bm{e}_\varphi\\
    &-(n+2)E\bm{e}_\varphi\bm{e}_r
    +(F-(n+1)Y)\bm{e}_\varphi\bm{e}_\theta
    +H\bm{e}_\varphi\bm{e}_\varphi
    \end{split}
\end{equation}

\begin{equation}
    \begin{split}
    &\sqrt{(n+1)(n+2)(2n+1)(2n+3)}\bm{Y}^m_{n,n+1,n+2}=
    (n+1)(n+2)Y\bm{e}_r\bm{e}_r
    -(n+2)E\bm{e}_r\bm{e}_\theta
    -(n+2)G\bm{e}_r\bm{e}_\varphi\\
   &-(n+2)E\bm{e}_\theta\bm{e}_r
    +(F-(n+1)Y)\bm{e}_\theta\bm{e}_\theta
    +H\bm{e}_\theta\bm{e}_\varphi
    -(n+2)G\bm{e}_\varphi\bm{e}_r
    +H\bm{e}_\varphi\bm{e}_\theta
    -(F+(n+1)^2Y)\bm{e}_\varphi\bm{e}_\varphi
    \end{split}
\end{equation}
\label{eq:rank-2-def}
\end{subequations}

Alternatively, rank $2$ tensors can be expanded using Zerilli tensors, $\bm{\mathrm{A}}=\sum_{n,m,l,n_2}A^{(l)}_{n,n_2;m}\bm{T}^{(l)}_{n,n_2;m}$,which are linearly related to rank 2 tensor spherical harmonics as
\begin{subequations}
\begin{equation}
    \bm{T}_{n,n_2;m}^{(l)}=\sum_{n_1}(-1)^{n+n_2+l}\Lambda(n_1,l)\begin{Bmatrix}
1 & l & 1\\
n & n_1 & n_2
\end{Bmatrix}\bm{Y}_{n,n_1,n_2}^m,
\end{equation}
\label{eq:Zerelli-transformation-0}
\end{subequations}
 The terms between curly brackets are Wigner's 6-j symbols. As Zerilli tensors are obtained by an unitary transformation from tensor spherical harmonics, it follows that they are also orthonormal with respect to the inner-product. We can transform between the components of a tensor in tensor spherical harmonics $A^m_{n,n_1,n_2}$, and in the Zerilli basis, $A^{(l)}_{n,n_2;m}$, as 
\begin{subequations}
\begin{equation}
   A^m_{n,n_1,n_2}=\sum_{l}(-1)^{n+n_2+l}\Lambda(n_1,l)\begin{Bmatrix}
1 & l & 1\\
n & n_1 & n_2
\end{Bmatrix}A_{n,n_2;m}^{(l)}
\end{equation}
\label{eq:Zerelli-to-spherical-harmonics}
\end{subequations}

\subsection{Operators}
The following operators are used
\begin{subequations}
\begin{equation}
    \partial_{n}^{n'}=\begin{dcases}
    \partial_r+\frac{n+1}{r},& \text{if } n=n'+1\\
    \partial_r-\frac{n}{r},& \text{if } n=n'-1
\end{dcases}
\end{equation}
\begin{equation}
G(n_k,n_{k+1})=(-1)^{n_k}\Lambda(n_{k+1})\begin{pmatrix}
n_k & n_{k+1} & 1\\
0 & 0 & 0
\end{pmatrix}=\frac{1}{\sqrt{2n_k+1}}\begin{dcases}
    \sqrt n_k,& \text{if } n_{k+1}=n_{k}-1\\
    -\sqrt{ n_k+1},& \text{if } n_{k+1}=n_{k}+1\\
    0,& \text{if } \text{otherwise }\\
\end{dcases}
\end{equation}
\label{eq:derivative-operators}
\end{subequations}
\begin{equation}
    D_n=\partial_r^2+\frac{2}{r}\partial_r-\frac{n(n+1)}{r^2}
    \label{eq:D_operator}
\end{equation}

\section{Tensor Spherical Harmonics Integrals}
\label{ap:coupling-coefficients}
To obtain both the coupling coefficients and the energy dissipation spectra the following integral should be evaluated
\begin{subequations}
\begin{equation}
    \left(\bm{T}^{(l_\alpha)}_{n_{\alpha},n_{2\alpha};m_\alpha}:\overline{\bm{T}^{(l_\beta)}_{n_\beta,n_{2\beta};m_\beta}}\cdot Y_{n_\nu}^{m_\nu}\right)=\frac{1}{4\pi}\int {\bm{T}^{(l_\alpha)}_{n_\alpha,n_{2\alpha};m_\alpha}} : \overline {\bm{T}^{(l_\beta)}_{n_\beta,n_{2\beta};m_\beta}}\ Y_{n_\nu}^{m_\nu}\text{d}\Omega,
\end{equation}

\begin{equation}
    \left(\bm{T}^{(l_\alpha)}_{n_{\alpha},n_{2\alpha};m_\alpha}:{\bm{T}^{(l_\beta)}_{n_\beta,n_{2\beta};m_\beta}}\cdot\overline{Y_{n_\nu}^{m_\nu}}\right)=\frac{1}{4\pi}\int {\bm{T}^{(l_\alpha)}_{n_\alpha,n_{2\alpha};m_\alpha}}: {\bm{T}^{(l_\beta)}_{n_\beta,n_{2\beta};m_\beta}}\overline{Y_{n_\nu}^{m_\nu}}\text{d}\Omega.
\end{equation}

\label{eq:spherical-harmonics-integrals-0}
\end{subequations}
Eq.~(\ref{eq:spherical-harmonics-integrals-0}a) is required to compute the coupling coefficients and (\ref{eq:spherical-harmonics-integrals-0}b) to compute the energy dissipation integrals. The two expressions can be obtained from 

\begin{equation}    
\left(\bm{T}^{(l_\alpha)}_{n_{\alpha},n_{2\alpha};m_\alpha}:{\bm{T}^{(l_\beta)}_{n_\beta,n_{2\beta};m_\beta}}\cdot Y_{n_\nu}^{m_\nu}\right)=\frac{1}{4\pi}\int {\bm{T}^{(l_\alpha)}_{n_\alpha,n_{2\alpha};m_\alpha}} : {\bm{T}^{(l_\beta)}_{n_\beta,n_{2\beta};m_\beta}}\ Y_{n_\nu}^{m_\nu}\text{d}\Omega,
\end{equation}
using the complex conjugate properties of tensor spherical harmonics (Eq.~(\ref{eq:complex-conjugate-tensor-spherical-harmonic})). To evaluate $\left(\bm{T}^{(l_\alpha)}_{n_{\alpha},n_{2\alpha};m_\alpha}:{\bm{T}^{(l_\beta)}_{n_\beta,n_{2\beta};m_\beta}}\cdot Y_{n_\nu}^{m_\nu}\right)$, we write it in terms of rank-2 spherical harmonics $\bm Y^m_{n(2)}$ (Eq.~\ref{eq:Zerelli-transformation-0}):

\begin{equation}
    \begin{split}
    &\left(\bm{T}^{(l_\alpha)}_{n_{\alpha},n_{2\alpha},m_\alpha}:{\bm{T}^{(l_\beta)}_{n_\beta,n_{2\beta};m_\beta}}\cdot Y_{n_\nu}^{m_\nu}\right)=(-1)^{n_{\alpha}+n_{2\alpha}+l_\alpha+n_\beta+n_{2\beta}+l_\beta}\\
    &\sum_{n_{1\alpha},n_{1\beta}}\Lambda(n_{1\alpha},n_{1\beta},l_{\alpha},l_\beta)
    \begin{Bmatrix}
    1 & l_\beta & 1\\
    n_{\beta}& n_{1\beta} & n_{2\beta}\\
    \end{Bmatrix}
     \begin{Bmatrix}
    1 & l_\alpha & 1\\
    n_\alpha & n_{1\alpha} & n_{2\alpha}\\
    \end{Bmatrix}
    \frac{1}{4\pi}\int \bm{Y}_{n_\alpha(2)}: \bm{Y}_{n_\beta(2)}Y_{n_\nu}^{m_\nu}\text{d}\Omega,
    \end{split}
\end{equation}
which can be obtained using the expressions provided in \cite{James1976} \S4

\begin{equation}
\begin{split}
   &\frac{1}{4\pi}\int \bm{Y}_{n_\alpha(2)} : \bm{Y}_{n_\beta(2)} Y_{n_\nu}^{m_\nu}\text{d}\Omega=\\
    &(-1)^{n_{2\beta}+n_{1\beta}+n_{1\alpha}+n_{\alpha}}\Lambda(n_{\alpha}(2),n_\beta(2),n_\nu)
    \begin{pmatrix}
    n_{2\alpha} & n_{2\beta} & n_\nu\\
     0 & 0 & 0\\
    \end{pmatrix}
    \begin{pmatrix}
    n_\alpha & n_\beta & n_\nu\\
     m_\alpha & m_\beta & m_\nu\\
    \end{pmatrix}
    \begin{Bmatrix}
    n_\alpha & n_{1\alpha} & 1\\
    n_{1\beta} & n_{\beta} & n_{\nu}\\
    \end{Bmatrix}
    \begin{Bmatrix}
    n_{1\alpha} & n_{2\alpha} & 1\\
    n_{2\beta} & n_{1\beta} & n_{\nu}\\
    \end{Bmatrix}
\end{split}
\end{equation}
We have used the abbreviation $\Lambda(n(k))=\sqrt{(2n+1)(2n_1+1)\hdots(2n_k+1)}$

The amount of integrals that need to be evaluated is greatly reduced by noting that many are  $0$.  $\left(\bm{T}^{(l_\alpha)}_{n_{\alpha},n_{2\alpha},m_\alpha}:{\bm{T}^{(l_\beta)}_{n_\beta,n_{2\beta};m_\beta}}\cdot Y_{n_\nu}^{m_\nu}\right)$ is $0$ unless

\begin{enumerate}
        \item $l_\alpha=l_\beta$
        \item $n_\alpha$, $n_\beta$ and $n_\nu$ satisfy the triangular inequality ($ \lvert n_\alpha-n_\beta \rvert\leq n_\nu\leq n_\alpha+n_\beta$)
        \item $n_{2\alpha}$, $n_{2\beta}$, and $n_{\nu}$ satisfy the triangular inequality 
        \item $n_{2\alpha}+n_{2\beta}+n_\nu$ is even
        \item If $m_\alpha=m_\beta=m_\nu$, $n_{\alpha}+n_{\beta}+n_\nu$ is even
        \item $m_\beta+m_\alpha+m_\nu$=0
       
\end{enumerate}

The previous set of properties can be used to deduce a set of selection rules for the excited modes. A mode of degree and orders $(n_0,m_0)$ together with lateral variations of degree and order $(n_1,m_1)$ results in modes of degree $(n_2,m_2)$ if: 
\begin{equation}
\begin{split}
   &(n_0,m_0)^x\bigotimes(n_1,m_1)\Rightarrow(n_2,m_2)^x\quad n_2\in\mathcal{N}_1\quad m_2=m_0+m_1\\
   &(n_0,m_0)^x\bigotimes(n_1,m_1)\Rightarrow(n_2,m_2)^y\quad n_2\in\mathcal{N}_2\quad m_2=m_0+m_1,\quad  \lvert m_0 \rvert+ \lvert m_1 \rvert+ \lvert m_2  \rvert\neq0\\
   &\mathcal{N}_1: n_2= \lvert n_0-n_1 \rvert+2i,\quad 0\leq i\leq\frac{1}{2}\left(n_0+n_1- \lvert n_0-n_1 \rvert\right)\\
   &\mathcal{N}_2: n_2= \lvert n_0-n_1 \rvert+2i+1,\quad 0\leq i\leq\frac{1}{2}\left(n_0+n_1- \lvert n_0-n_1 \rvert\right)-1
 \end{split}
\end{equation}
where $x$ or $y$ indicate if the mode is spheroidal or toroidal and the cases  $x=s$, $y=t$, and $x=t$, $y=s$ can be considered.

\section{Explicit Form of the Equations}
\label{ap:explicit-equations}
The components of the strain tensor follow from the gradient of the displacement vector  (Eq.~(\ref{eq:strain-trace-free-sym}))
\begin{subequations}
\begin{equation}
\begin{split}
    \epsilon_{n,n;m}^{(0)}=-\frac{1}{\sqrt{3}}\chi_n^m,
\end{split}
\end{equation}

\begin{equation}
\begin{split}
    \epsilon_{n,n-2;m}^{(2)}=\sqrt{\frac{n-1}{2n-1}}\left(\partial_r u^m_{n,n-1}+\frac{n}{r}u^m_{n,n-1}\right),
\end{split}
\end{equation}

\begin{equation}
\begin{split}
    \epsilon_{n,n-1;m}^{(2)}=\frac{1}{\sqrt{2}}\sqrt{\frac{n-1}{2n+1}}\left(\partial_r u^m_{n,n}+\frac{n+1}{r}u^m_{n,n}\right),
\end{split}
\end{equation}

\begin{equation}
\begin{split}
    \epsilon_{n,n;m}^{(2)}=&-\sqrt{\frac{(2n+3)(2n+2)}{12(2n-1)(2n+1)}}\left(\partial_r-\frac{n-1}{r}\right)u_{n,n-1,n}^m\\
    &+\sqrt{\frac{n(2n-1)(n+1)}{3(2n+3)(2n+2)(2n+1)}}\left(\partial_r+\frac{n+2}{r}\right)u_{n,n+1,n}^m
\end{split},
\end{equation}

\begin{equation}
\begin{split}
    \epsilon_{n,n+1;m}^{(2)}=-\frac{1}{\sqrt{2}}\sqrt{\frac{n+2}{2n+1}}\left(\partial_r u^m_{n,n}-\frac{n}{r}u^m_{n,n}\right),
\end{split}
\end{equation}

\begin{equation}
\begin{split}
    \epsilon_{n,n+2;m}^{(2)}=-\sqrt{\frac{n+2}{2n+3}}\left(\partial_r u^m_{n,n+1}-\frac{n+1}{r}u^m_{n,n+1}\right).
\end{split}
\end{equation}
\label{eq:strain-tensor-explicit}
\end{subequations}
$\chi_n^m$ is the divergence of the displacement vector
\begin{equation}
    \begin{split}
    \chi_n^m=&\frac{1}{\sqrt{2n+1}}\left[\sqrt{n}\partial_ru_{n,n-1}^m-\sqrt{n+1}\partial_ru_{n,n+1}^m-\frac{1}{r}\left((n-1)\sqrt{n}u_{n,n-1}^m+\sqrt{n+1}(n+2)u_{n,n+1}^m\right)\right]\\
    &=\partial_r U_n^m+\frac{2U_n^m}{r}-\frac{n(n+1)}{r}V_n^m.
    \end{split}
\end{equation}

Eq.~(\ref{eq:strain-tensor-explicit}) can be written more  concisely as
\begin{equation}
\bm{\epsilon}_{n,n_2,m}^{(l)}=\left(\bm{\mathrm{E}}_{\partial_r}\partial_r+\frac{\bm{\mathrm{E}}}{r}\right)\bm{{u}}_{n,n_1}^{m}
\end{equation}
with $\bm{\epsilon}_{n,n_2,m}^{(l)}$ being a vector containing the elements of the strain tensor, $\bm{\epsilon}_{n,n_2,m}^{(l)}=(\hdots\epsilon_{n,n,m}^{(0)},\epsilon_{n,n-2,m}^{(2)} \epsilon_{n,n-1,m}^{(2)}, \epsilon_{n,n,m}^{(2)}, \epsilon_{n,n+1,m}^{(2)}, \epsilon_{n,n+2,m}^{(2)} \hdots)^\dag$.

The components of the stress tensor can be written in terms of the strain tensor and the material properties 
\begin{subequations}
    
\begin{equation}
    \sigma_{n,n;m}^{(0)}=3\kappa_0\epsilon_{n,n;m}^{(0)}    +3\kappa_0\sum_{\substack{n_\beta,m_\beta,\\n_\alpha,m_\alpha}}\kappa_{n_\beta}^{m_\beta}\epsilon_{n_\alpha,n_\alpha;m_\alpha}^{(0)}\left(\bm{T}^{(0)}_{n_{\alpha},n_{\alpha};m_\alpha}:\overline{\bm{T}^{(0)}_{n,n;m}}\cdot Y_{n_\beta}^{m_\beta}\right),
\end{equation}

\begin{equation}
    \sigma_{n,n_2;m}^{(2)}=2\hat\mu_0\epsilon_{n,n_2;m}^{(2)}    +2\mu_0\sum_{\substack{n_\beta,m_\beta,\\n_\alpha,m_\alpha,n_{2\alpha}}}\hat\mu_{n_\beta}^{m_{\beta}}\epsilon_{n_\alpha,n_{2\alpha};m_\alpha}^{(2)}\left(\bm{T}^{(2)}_{n_{\alpha},n_{2\alpha};m_\alpha}:\overline{\bm{T}^{(2)}_{n,n_{2};m}}\cdot Y_{n_\beta}^{m_\beta}\right),
\end{equation}
\label{eq:stress-tensor-explicit}
\end{subequations}
, or in matrix form
\begin{equation}
    \bm\sigma_{n,n_2,m}^{(l)}=(\bm{\mathrm{R}}_u+\bm{\mathrm{R}}_v)\bm\epsilon_{n,n_2,m}^{(l)}    
\end{equation}
with $\bm{\sigma}_{n,n_2,m}^{(l)}$ being a vector containing the components of the stress tensor, $\bm{\sigma}_{n,n_2,m}^{(l)}=(\hdots\sigma_{n,n,m}^{(0)},\sigma_{n,n-2,m}^{(2)} \sigma_{n,n-1,m}^{(2)}, \sigma_{n,n,m}^{(2)}, \sigma_{n,n+1,m}^{(2)}, \sigma_{n,n+2,m}^{(2)} \hdots)^\dag$; $\bm{\mathrm{R}_u}$ a matrix with the mean mechanical properties of the material,

\begin{subequations}
    \begin{equation}
     \matrixB{R}_u=\begin{bmatrix}
    \left(\matrixB{R}_u\right)_{0,0} &  & \\ 
    & \ddots & \\ 
    &  &  \left(\matrixB{R}_u\right)_{\infty,\infty}
    \end{bmatrix}        
    \end{equation}

    \begin{equation}      \left(\matrixB{R}_u\right)_{n,m}=\mathrm{diag}\left(3\kappa_0,2\mu_0,2\mu_0,2\mu_0,2\mu_0,2\mu_0\right)   
\end{equation}
\end{subequations}

and $\bm{\mathrm{R}}_v$ a matrix that includes both the lateral variations of mechanical properties of the material and the coupling coefficients.

\begin{subequations}
    \begin{equation}
     \matrixB{R}_v=
   \begin{bmatrix}
   \left(\matrixB{R}_v\right)_{0,0}^{0,0} & \cdots & \left(\matrixB{R}_v\right)^{\infty,\infty}_{0,0} \\
   \vdots & \ddots & \vdots \\
   \left(\matrixB{R}_v\right)^{0,0}_{0,0} & \cdots & \left(\matrixB{R}_v\right)_{\infty,\infty}^{0,0} \\
 \end{bmatrix}      
    \end{equation}

    \begin{equation}
    \left(\matrixB{R}_v\right)_{n,m}^{n_\alpha,m_\alpha}=  \sum_{\substack{n_\beta,m_\beta}}\matrixB{R}_{n_\beta,m_\beta}\underset{n_\beta,m_\beta}{\matrixB{C}}_{n,m}^{n_\alpha,m_\alpha}
    \end{equation}
    
\begin{equation}      \matrixB{R}_{n_\beta,m_\beta}=\mathrm{diag}\left(3\kappa_0\kappa_{n_\beta}^{m_\beta},2\mu_0\mu_{n_\beta}^{m_\beta},2\mu_0\mu_{n_\beta}^{m_\beta},2\mu_0\mu_{n_\beta}^{m_\beta},2\mu_0\mu_{n_\beta}^{m_\beta},2\mu_0\mu_{n_\beta}^{m_\beta}\right)
\end{equation}

    \begin{equation}
   \underset{n_\beta,m_\beta}{\matrixB{C}}_{n,m}^{n_\alpha,m_\alpha}=   
   \begin{bmatrix}
   {\underset{n_\beta,m_\beta}{C^{(0)}}}_{n,n,m}^{n_\alpha,n_\alpha,m_\alpha} & 0 & 0 & 0 & 0 & 0 \\
   0&
   {\underset{n_\beta,m_\beta}{C^{(2)}}}_{n,n-2,m}^{n_\alpha,n_\alpha-2,m_\alpha}&
   {\underset{n_\beta,m_\beta}{C^{(2)}}}_{n,n-2,m}^{n_\alpha,n_\alpha-1,m_\alpha} & 
   {\underset{n_\beta,m_\beta}{C^{(2)}}}_{n,n-2,m}^{n_\alpha,n_\alpha,m_\alpha} &
   {\underset{n_\beta,m_\beta}{C^{(2)}}}_{n,n-2,m}^{n_\alpha,n_\alpha+1,m_\alpha} & 
   {\underset{n_\beta,m_\beta}{C^{(2)}}}_{n,n-2,m}^{n_\alpha,n_\alpha+2,m_\alpha} & \\
   0&
   {\underset{n_\beta,m_\beta}{C^{(2)}}}_{n,n-1,m}^{n_\alpha,n_\alpha-2,m_\alpha}&
   {\underset{n_\beta,m_\beta}{C^{(2)}}}_{n,n-1,m}^{n_\alpha,n_\alpha-1,m_\alpha} & 
   {\underset{n_\beta,m_\beta}{C^{(2)}}}_{n,n-1,m}^{n_\alpha,n_\alpha,m_\alpha} &
   {\underset{n_\beta,m_\beta}{C^{(2)}}}_{n,n-1,m}^{n_\alpha,n_\alpha+1,m_\alpha} & 
   {\underset{n_\beta,m_\beta}{C^{(2)}}}_{n,n-1,m}^{n_\alpha,n_\alpha+2,m_\alpha} & \\
   0&
   {\underset{n_\beta,m_\beta}{C^{(2)}}}_{n,n,m}^{n_\alpha,n_\alpha-2,m_\alpha}&
   {\underset{n_\beta,m_\beta}{C^{(2)}}}_{n,n,m}^{n_\alpha,n_\alpha-1,m_\alpha} & 
   {\underset{n_\beta,m_\beta}{C^{(2)}}}_{n,n,m}^{n_\alpha,n_\alpha,m_\alpha} &
   {\underset{n_\beta,m_\beta}{C^{(2)}}}_{n,n,m}^{n_\alpha,n_\alpha+1,m_\alpha} & 
   {\underset{n_\beta,m_\beta}{C^{(2)}}}_{n,n,m}^{n_\alpha,n_\alpha+2,m_\alpha} & \\
   0&
   {\underset{n_\beta,m_\beta}{C^{(2)}}}_{n,n+1,m}^{n_\alpha,n_\alpha-2,m_\alpha}&
   {\underset{n_\beta,m_\beta}{C^{(2)}}}_{n,n+1,m}^{n_\alpha,n_\alpha-1,m_\alpha} & 
   {\underset{n_\beta,m_\beta}{C^{(2)}}}_{n,n+1,m}^{n_\alpha,n_\alpha,m_\alpha} &
   {\underset{n_\beta,m_\beta}{C^{(2)}}}_{n,n+1,m}^{n_\alpha,n_\alpha+1,m_\alpha} & 
   {\underset{n_\beta,m_\beta}{C^{(2)}}}_{n,n+1,m}^{n_\alpha,n_\alpha+2,m_\alpha} & \\
    0&
   {\underset{n_\beta,m_\beta}{C^{(2)}}}_{n,n+2,m}^{n_\alpha,n_\alpha-2,m_\alpha}&
   {\underset{n_\beta,m_\beta}{C^{(2)}}}_{n,n+2,m}^{n_\alpha,n_\alpha-1,m_\alpha} & 
   {\underset{n_\beta,m_\beta}{C^{(2)}}}_{n,n+2,m}^{n_\alpha,n_\alpha,m_\alpha} &
   {\underset{n_\beta,m_\beta}{C^{(2)}}}_{n,n+2,m}^{n_\alpha,n_\alpha+1,m_\alpha} & 
   {\underset{n_\beta,m_\beta}{C^{(2)}}}_{n,n+2,m}^{n_\alpha,n_\alpha+2,m_\alpha} & \\
 \end{bmatrix}  
\end{equation}
\label{eq:stress-strain-matrix}
\end{subequations}

where we have introduced the notation $\underset{n_\beta,m_\beta}{C^{(l)}}_{n,n_2,m}^{n_\alpha,n_2\alpha,m_\alpha}=\left(\bm{T}^{(l)}_{n_{\alpha},n_{2\alpha};m_\alpha}:\overline{\bm{T}^{(l)}_{n,n_{2};m}}\cdot Y_{n_\beta}^{m_\beta}\right)$. $\matrixB{R}_{n_\beta,m_\beta}$ contains the mechanical properties and $\underset{n_\beta,m_\beta}{\matrixB{C}}_{n,m}^{n_\alpha,m_\alpha}$ is a matrix with the coupling coefficients. If there are no lateral variations, $\bm{\mathrm{R}}_v$ is $0$.

The explicit form  of the momentum (Eq.~\ref{eq:projected-momentum-equation}) and Poisson equations (Eq.~\ref{eq:projected-Poisson}) are 
\begin{subequations}
\begin{equation}
    \begin{split}
    &-\sqrt{\frac{n}{6n+3}}\left(\partial_r+\frac{n+1}{r}\right)\sigma^{(0)}_{n,n;m}-\sqrt{\frac{(2n+3)(2n+2)}{12(2n-1)(2n+1)}}\left(\partial_r+\frac{n+1}{r}\right)\sigma^{(2)}_{n,n;m}\\&+\sqrt{\frac{n-1}{2n-1}}\left(\partial_r-\frac{n-2}{r}\right)\sigma^{(2)}_{n,n-2;m}
    -\rho_0\frac{n}{2n+1}\left(\partial_r g+\frac{2n}{r}g\right)u^m_{n,n-1}+\rho_0\frac{\sqrt{n(n+1)}}{2n+1}\left(\partial_r g-\frac{g}{r}\right) u^m_{n,n+1}\\
    &-\rho_0\sqrt{\frac{n}{2n+1}}\left(\partial_r+\frac{n+1}{r}\right)\phi^m_n=0
    \end{split}
\end{equation}
\begin{equation}
    \begin{split}
    &\sqrt{\frac{n+1}{6n+3}}\left(\partial_r-\frac{n}{r}\right)\sigma^{(0)}_{n,n;m}+\sqrt{\frac{n(n+1)(2n-1)}{3(2n+1)(2n+2)(2n+3)}}\left(\partial_r-\frac{n}{r}\right)\sigma^{(2)}_{n,n;m}-\sqrt{\frac{n+2}{2n+3}}\left(\partial_r+\frac{n+3}{r}\right)\sigma^{(2)}_{n,n+2;m}\\
    &\rho_0\frac{\sqrt{n(n+1)}}{2n+1}\left(\partial_r g-\frac{g}{r}\right)u^m_{n,n-1}+\rho_0\frac{n+1}{2n+1}\left(-\partial_r g+g\frac{2n+2}{r}\right) u^m_{n,n+1}+\rho_0\sqrt{\frac{n+1}{2n+1}}\left(\partial_r-\frac{n}{r}\right)\phi^m_n=0
    \end{split}
\end{equation}
\begin{equation}
    \sqrt{n-1}\left(\partial_r-\frac{n-1}{r}\right)\sigma_{n,n-1;m}^{(2)}-\sqrt{n+2}\left(\partial_r+\frac{n+2}{r}\right)\sigma_{n,n+1;m}^{(2)}=0
\end{equation}

\begin{equation}
    \begin{split}
    \left(\partial_r^2+\frac{2\partial_r}{r}-\frac{n(n+1)}{r^2}\right)\phi_n^m=
    -4\pi G \Bigg[&\sqrt{\frac{n}{2n+1}}\left(\rho_0\partial_r+\partial_r\rho_0-\rho_0\frac{n-1}{r}\right)u^m_{n,n-1}-\\
    &\sqrt{\frac{n+1}{2n+1}}\left(\rho_0\partial_r+\partial_r\rho_0+\rho_0\frac{n+2}{r}\right)u^m_{n,n+1}\Bigg]
    \end{split}
\end{equation}
\label{eq:expanded-equations}
\end{subequations}

To cast the governing equations in terms $U_n^m,V_n^m,W_n^m,R_n^m,S_n^m,T_n^m,\phi_n^m,\partial_r \phi_n^m$ and their derivatives, we first need to obtain the radial components of the stress tensor \cite[][Eq. 3.22]{James1976}

\begin{subequations}
\begin{equation}
\begin{split}
    R_n^m=&
    -\frac{1}{\sqrt{3}}\sigma^{(0)}_{n,n;m}
    -\frac{\sqrt{n}}{2n+1}\left(\sqrt{\frac{(2n+3)(2n+2)}{12(2n-1)}}+\sqrt{\frac{(2n-1)(n+1)^2}{3(2n+2)(2n+3)}}\right)\sigma^{(2)}_{n,n;m}\\
    &+\sqrt{\frac{n(n-1)}{(2n-1)(2n+1)}}\sigma^{(2)}_{n,n-2;m}
    +\sqrt{\frac{(n+1)(n+2)}{(2n+1)(2n+3)}}\sigma^{(2)}_{n,n+2;m}=\matrixB{C}_{\sigma\rightarrow R_n^m}\bm{\sigma}_{n,n_2,m}^{(l)}
\end{split}
\end{equation}
\begin{equation}
    \begin{split}
    S_n^m=&\frac{1}{\sqrt{2n+1}}\Bigg[
    \left(-\sqrt{\frac{(2n+3)(2n+2)}{(12n(2n-1)(2n+1))}}+\sqrt{\frac{n(2n-1)}{3(2n+1)(2n+2)(2n+3)}}\right)\sigma_{n,n;m}^{(2)}\\
    &+\sqrt{\frac{n-1}{(2n-1)n}}\sigma^{(2)}_{n,n-2;m}
    -\sqrt{\frac{n+2}{(2n+3)(n+1)}}\sigma^{(2)}_{n,n+2;m}
    \Bigg]=\matrixB{C}_{\sigma\rightarrow S_n^m}\bm{\sigma}_{n,n_2,m}^{(l)}
    \end{split}
\end{equation}
\begin{equation}
    T_n^m=\frac{\iu}{\sqrt{2n(n+1)(2n+1)}}\left[
    \sqrt{n-1}\sigma^{(2)}_{n,n-1;m}
    -\sqrt{n+2}\sigma^{(2)}_{n,n+1;m}
    \right]=\matrixB{C}_{\sigma\rightarrow T_n^m}\bm{\sigma}_{n,n_2,m}^{(l)}
\end{equation}
\label{eq:stress-radial-expanded}
\end{subequations}
and then project the momentum equation to the scaloidal-poloidal-toroidal basis
\begin{subequations}
\begin{equation}
\begin{split}
       -\partial_r R_n^m=&\frac{1}{2n+1}\left[-(n+1)\sqrt{\frac{(2n+3)(2n+2)n}{12(2n-1)}}+n\sqrt{\frac{n(n+1)^2(2n-1)}{3(2n+2)(2n+3)}}\right]\frac{\sigma^{(2)}_{n,n;m}}{r}\\
       &-(n-2)\sqrt{\frac{n(n-1)}{(2n-1)(2n+1)}}\frac{\sigma^{(2)}_{n,n-2;m}}{r}+
       (n+3)\sqrt{\frac{(n+1)(n+2)}{(2n+3)(2n+1)}}\frac{\sigma^{(2)}_{n,n+2;m}}{r}\\
       &-\rho_0\partial_r\phi_n^m+\rho g\chi_n^m-\rho\partial_r(gU_n^m)=\frac{\matrixB{C}_{\sigma\rightarrow \partial_r R_n^m}}{r}\bm{\sigma}_{n,n_2,m}^{(l)}-\rho_0\partial_r\phi_n^m+\rho g\chi_n^m-\rho\partial_r(gU_n^m)
\end{split}
\end{equation}
\begin{equation}
\begin{split}
       -\partial_r S_n^m=&-\frac{1}{\sqrt{3}}\frac{\sigma^{(0)}_{n,n;m}}{r}-
       \frac{1}{2n+1}\left[(n+1)\sqrt{\frac{(2n+3)(2n+2)}{12n(2n-1)}}+n\sqrt{\frac{n(2n-1)}{3(2n+2)(2n+3)}}\right]\frac{\sigma^{(2)}_{n,n;m}}{r}\\
       &-(n-2)\sqrt{\frac{(n-1)}{(2n-1)(2n+1)n}}\frac{\sigma^{(2)}_{n,n-2;m}}{r}-
       (n+3)\sqrt{\frac{(n+2)}{(2n+3)(2n+1)(n+1)}}\frac{\sigma^{(2)}_{n,n+2;m}}{r}\\
       &-\frac{\rho}{r}\phi_n^m-\frac{\rho g}{r}U_n^m=\frac{\matrixB{C}_{\sigma\rightarrow \partial_r S_n^m}}{r}\bm{\sigma}_{n,n_2,m}^{(l)}-\frac{\rho}{r}\phi_n^m-\frac{\rho g}{r}U_n^m   
\end{split}
\end{equation}
\begin{equation}
    \partial_r T_n^m=\frac{\iu}{r\sqrt{2n(n+1)(2n+1)}}\left[(n-1)^{3/2}\sigma^{(2)}_{n,n-1;m}+(n+2)^{3/2}\sigma^{(2)}_{n,n+1;m}\right]=\frac{\matrixB{C}_{\sigma\rightarrow \partial_r T_n^m}}{r}\bm{\sigma}_{n,n_2,m}^{(l)}
\end{equation}
\label{eq:expanded-momentum}
\end{subequations}
For conciseness, we have introduced the matrices $\matrixB{C}_{\sigma\rightarrow R_n^m}$, $\matrixB{C}_{\sigma\rightarrow S_n^m}$, $\matrixB{C}_{\sigma\rightarrow T_n^m}$ that allow to obtain the radial, solenoidal and toroidal components of the radial component of the stress tensors in terms of $\bm\sigma_{n,n_2,m}^{(l)}$; and $\matrixB{C}_{\sigma\rightarrow \partial r R_n^m}$, $\matrixB{C}_{\sigma\rightarrow \partial_r S_n^m}$, $\matrixB{C}_{\sigma\rightarrow \partial rT_n^n}$ that can be used to write  part of $\partial_r R_n^m$, $\partial_r S_n^m$ and $\partial_r T_n^m$  in terms of $\bm\sigma_{n,n_2,m}^{(l)}$.

Plugging the expressions for the stress tensor (Eq.~(\ref{eq:stress-tensor-explicit})) into Eqs.~(\ref{eq:stress-radial-expanded}) and (\ref{eq:expanded-momentum}),using the linear transformation between $U_n^m,V_n^m,W_n^m$ and $u_{n,n-1}^m,u_{n,n}^m,u_{n,n+1}^m$ (Eq.~(\ref{eq:displacement-vector-rad-tan})) for the resulting equations and Poisson's equation (Eq.~(\ref{eq:expanded-equations})d), we find

\begin{subequations}
\begin{equation}
    \begin{split}
    -\rho_0\partial_r\phi_n^m-\rho_0\partial_r(gU_n^m)+\rho_0 g\chi_n^m+\partial_r R_n^m+
    \frac{\hat\mu_0}{r^2}\left[4r\partial_r U_n^m-4U_n^m+n(n+1)(3V_n^m-U_n^m-r\partial_r V_n^m)\right]+\\
    \frac{\matrixB{C}_{\sigma\rightarrow\partial_r R_n^m}}{r}\matrixB{R}_v\left(\matrixB{E}_{\partial_r}\partial_r+\frac{\matrixB{E}}{r}\right)\matrixB{C}_{U\rightarrow u}\bm{U}_n^m=0
    \end{split}
\end{equation}
\begin{equation}
    \begin{split}
    -\frac{\rho_0}{r}\phi^m_n-\frac{\rho_0}{r}gU_n^m+\frac{\hat\lambda_0}{r}\chi_n^m+\partial_r S_n^m+\frac{\hat\mu_0}{r^2}\left[5U_n^m+3r\partial_rV_n^m-V_n^m-2n(n+1)V_n^m\right]+\\
    \frac{\matrixB{C}_{\sigma\rightarrow\partial_r S_n^m}}{r}\matrixB{R}_v\left(\matrixB{E}_{\partial_r}\partial_r+\frac{\matrixB{E}}{r}\right)\matrixB{C}_{U\rightarrow u}\bm{U}_n^m=0
    \end{split}
\end{equation}
\begin{equation}
    \partial_r T_{n}^m+\frac{3}{r}T_n^m-\frac{\hat\mu_0}{r^2}(n(n+1)-2)W_n^m+\frac{\matrixB{C}_{\sigma\rightarrow\partial_r T_n^m}}{r}\matrixB{R}_v\left(\matrixB{E}_{\partial_r}\partial_r+\frac{\matrixB{E}}{r}\right)\matrixB{C}_{U\rightarrow u}\bm{U}_n^m=0
\end{equation}
\begin{equation}
    \left(\partial_r^2+\partial_r\frac{2}{r}-\frac{n(n+1)}{r^2}\right)\phi_n^m=
    -4\pi G\left(\rho\chi_n^m+U_n^m\partial_r\rho_0\right)
\end{equation}
\begin{equation}
    R_n^m=\hat\lambda_0\chi_n^m+2\hat\mu_0\partial_rU_n^m
    +\matrixB{C}_{\sigma\rightarrow R_n^m}\matrixB{R}_v\left(\matrixB{E}_{\partial_r}\partial_r+\frac{\matrixB{E}}{r}\right)\matrixB{C}_{U\rightarrow u}\bm{U}_n^m
\end{equation}
\begin{equation}
    S_n^m=\hat\mu_0\left(\partial_r V_n^m+\frac{U_n^m-V_n^m}{r}\right)
    +\matrixB{C}_{\sigma\rightarrow S_n^m}\matrixB{R}_v\left(\matrixB{E}_{\partial_r}\partial_r+\frac{\matrixB{E}}{r}\right)\matrixB{C}_{U\rightarrow u}\bm{U}_n^m
\end{equation}
\begin{equation}
    T_n^m=\hat\mu_0\left[\partial_r W_n^m-\frac{W_n^m}{r}\right]
    +\matrixB{C}_{\sigma\rightarrow T_n^n}\matrixB{R}_v\left(\matrixB{E}_{\partial_r}\partial_r+\frac{\matrixB{E}}{r}\right)\matrixB{C}_{U\rightarrow u}\bm{U}_n^m
\end{equation}
\label{eq:final-explicit}
\end{subequations}

The terms with $\bm{\mathrm{R}}_v$ arise from lateral variations and couple equations involving different modes (see Eq.~\ref{eq:stress-strain-matrix}). Without lateral variations, Eq.~(\ref{eq:final-explicit})   reduces to the equations typical of a spherically-symmetric body. Using linear algebra, the previous set of equations can be cast as $\partial_r\bm{y} =\matrixB{D}\bm{y}$.

\section{Tidal Potential for a Synchronous Satellite}
\label{sec:tidal-potential-synch-sat}

The tidal potential experienced by a synchronous body in an eccentric orbit of period $T=2\pi/\omega$ is given by \citep[e.g.,][]{JaraOrue2011}
\begin{equation}
\begin{split}
  \phi^{T}=\left({{\phi}^{(T)}}_{2}^{0,+}Y_{2}^0+{{\phi^{(T)}}}_{2}^{2,+}Y_{2}^2+{{\phi^{(T)}}}_{2}^{-2,+}Y_{2}^{-2}\right)\ex^{\iu\omega t}+
  \left({{\phi^{(T)}}}_{2}^{0,-}Y_{2}^0+{{\phi^{(T)}}}_{2}^{-2,-}Y_{2}^{-2}+{{\phi^{(T)}}}_{2}^{2,-}Y_{2}^{2}\right)\ex^{-\iu\omega t}+O(e^2)
\end{split}
  \label{eq:tidal-potential}
\end{equation}
The ${{\phi^{(T)}}}_{2}^{0,+}$, ${{\phi^{(T)}}}_{2}^{0,-}$; ${{\phi^{(T)}}}_{2}^{2,+}$, ${{\phi^{(T)}}}_{2}^{2,-}$; and ${{\phi^{(T)}}}_{2}^{-2,+}$, ${{\phi^{(T)}}}_{2}^{-2,-}$  correspond to standing, westward propagating and eastward propagating waves, respectively. As the tidal potential is real, we have that ${\phi^{(T)}}_n^{-m,-}=(-1)^{m}{\phi^{(T)}}_n^{m,+}$. The amplitude of the different components of the tide are given in Table \ref{tab:tidal-comonents}.

\begin{table}
\begin{center}
\begin{tabular}{ c c}
 \hline
Tidal Component & Amplitude \\
\hline
 ${\phi^{(T)}}_{2}^{0,+}$ & $  \frac{3}{4}\sqrt{\frac{1}{5}}(\omega R)^2 e$ \\ 
 ${\phi^{(T)}}_{2}^{2,+}$ & $ \frac{1}{8}\sqrt{\frac{6}{5}}(\omega R)^2 e$  \\  
 ${\phi^{(T)}}_{2}^{-2,+}$ & $-\frac{7}{8}\sqrt{\frac{6}{5}}(\omega R)^2 e$ \\  \hline
\end{tabular}
\end{center}
\caption{Amplitude of the components of the eccentricity tide}
\label{tab:tidal-comonents}
\end{table}

\section{Test on the total energy dissipation}
\label{ap:energy-dissipation-test}
\begin{figure}[ht!]
\centering
\includegraphics[width=0.5\textwidth]{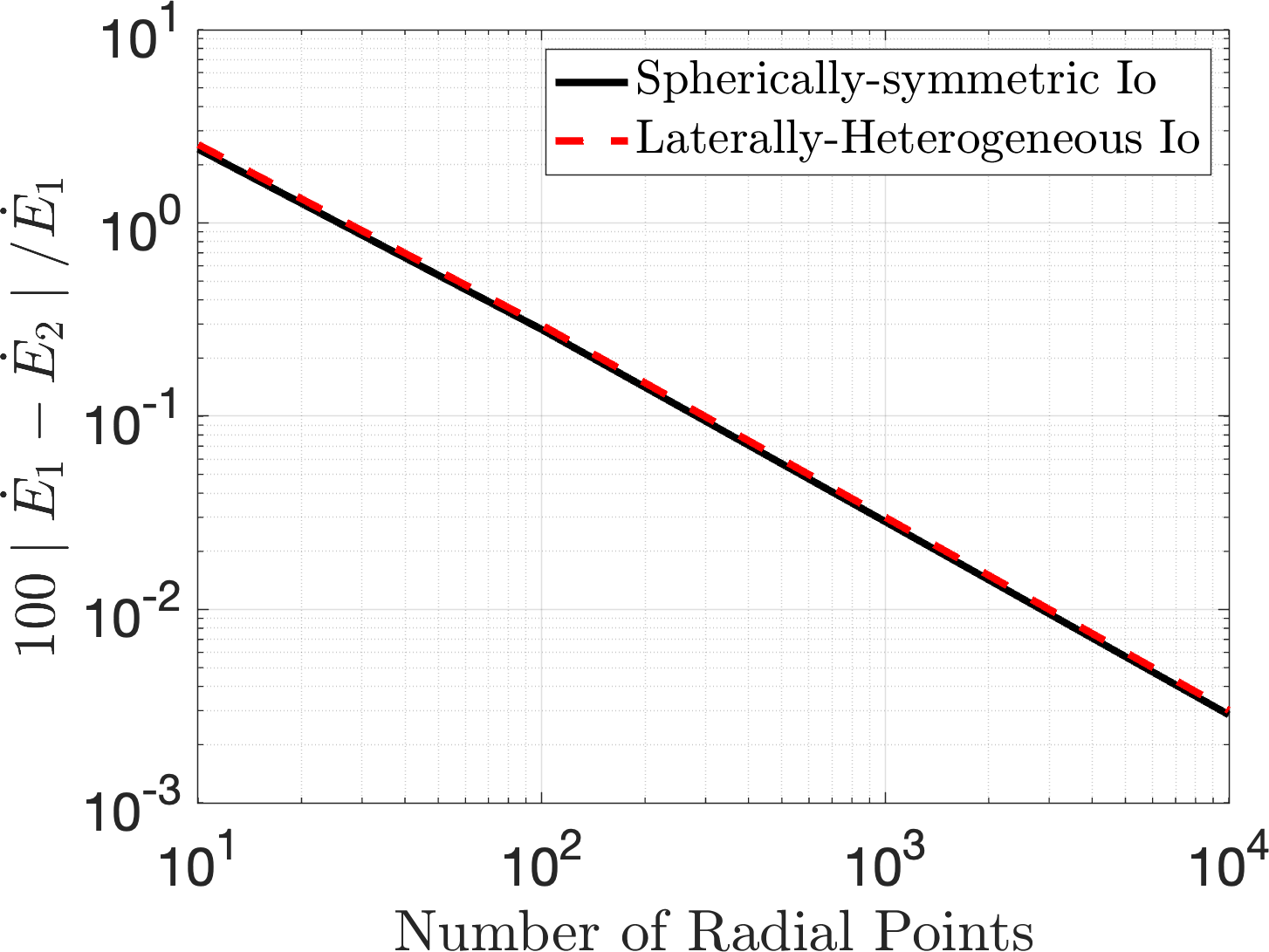}
\caption{Difference in tidal heating computed using the stress and strain tensors (Eq.~(\ref{eq:tidal-dissipation-0}), $\dot{E}_1$) and the work done by the tidal force (Eq.~(\ref{eq:energy-method-2}), $\dot{E}_2$). We assume the viscoelastic Io model of Table \ref{tab:non-dimensional-numbers} and zonal degree $2$ lateral viscosity variations with peak to peak amplitude of $50\%$ the mean viscosity.} 
\label{fig:test-energy}
\end{figure}

We test that the total energy dissipation obtained using the stress and strain tensors (Eqs.~(\ref{eq:tidal-dissipation-0})--(\ref{eq:energy-method-1})) and considering the work done by the tidal potential (Eqs.~(\ref{eq:energy-method-2})--(\ref{eq:energy-method-2-2})) is the same. We consider a viscoelastic Io that is spherically-symmetric and one that has zonal degree $2$ lateral viscosity variations with peak to peak amplitude of $50\%$ the mean viscosity and compute tidal heating using Eq.~(\ref{eq:energy-method-1}) ($\dot{E}_1$) and (\ref{eq:energy-method-2-2}) ($\dot{E}_2$). We find excellent agreement between the total energy dissipation obtained with the two methods, with the difference decreasing as the number of radial points used for the radial integration increases (Figure \ref{fig:test-energy}). Equivalent results are obtained for lateral variations of different amplitude, and degree and order. 

\bibliography{biblio_abb}{}
\bibliographystyle{aasjournal}

\end{document}